\documentclass{pasj00}

\SetRunningHead{Totani et al.}{SN Ia Delay Time Distribution}

\begin{document}
\title{Delay Time Distribution Measurement of Type Ia Supernovae by
  the Subaru/XMM-Newton Deep Survey and Implications for the
  Progenitor }

\author{
Tomonori \textsc{Totani}\altaffilmark{1},
Tomoki \textsc{Morokuma}\altaffilmark{2},
Takeshi \textsc{Oda}\altaffilmark{1},
Mamoru \textsc{Doi}\altaffilmark{3}, and
Naoki \textsc{Yasuda}\altaffilmark{4}
}

\altaffiltext{1}{Department of Astronomy, School of Science,
Kyoto University, Sakyo-ku, Kyoto 606-8502}

\altaffiltext{2}{National Astronomical Observatory, 2-21-1
Osawa, Mitaka, Tokyo, 181-8588}

\altaffiltext{3}{Institute of Astronomy, School of Science,
The University of Tokyo, 2-21-1 Osawa, Mitaka, Tokyo, 181-0015}

\altaffiltext{4}{Institute for Cosmic Ray Research, The University
of Tokyo, Kashiwa, Chiba, 227-8582}


\maketitle

\begin{abstract}
  The delay time distribution (DTD) of type Ia supernovae (SNe Ia)
  from star formation is an important clue to reveal the still unknown
  progenitor system of SNe Ia. Here we report on a measurement of the
  SN Ia DTD in a delay time range of $t_{\rm Ia} = $ 0.1--8.0 Gyr by
  using the faint variable objects detected in the Subaru/XMM-Newton
  Deep Survey (SXDS) down to $i' \sim 25.5$.  We select 65 SN
  candidates showing significant spatial offset from nuclei of the
  host galaxies having old stellar population at $z \sim$ 0.4--1.2,
  out of more than 1,000 SXDS variable objects.  Although
  spectroscopic type classification is not available for these, we
  quantitatively demonstrate that more than $\sim$80 \% of these
  should be SNe Ia. The DTD is derived using the stellar age estimates
  of the old galaxies based on 9 band photometries from optical to
  mid-infrared wavelength. Combined with the observed SN Ia rate in
  elliptical galaxies at the local universe, the DTD in $t_{\rm Ia}
  \sim $ 0.1--10 Gyr is well described by a featureless power-law as
  $f_D(t_{\rm Ia}) \propto t_{\rm Ia}^\alpha$ with $\alpha \sim
  -1$. The derived DTD is in excellent agreement with the generic
  prediction of the double-degenerate scenario, giving a strong
  support to this scenario. In the single-degenerate (SD) scenario,
  although predictions by simple analytic formulations have broad DTD
  shapes that are similar to the observation, DTD shapes calculated by
  more detailed binary population synthesis tend to have strong peaks
  at characteristic time scales, which do not fit the
  observation. This result thus indicates either that the SD channel
  is not the major contributor to SNe Ia in old stellar population, or
  that improvement of binary population synthesis theory is required.
  Various sources of systematic uncertainties are examined and tested,
  but our main conclusions are not affected significantly.
\end{abstract}

\section{Introduction}

It is widely believed that type Ia supernovae (SNe Ia) are
thermonuclear explosions of carbon-oxygen white dwarfs in binary
systems, triggered when a white dwarf grows up to the Chandrasekhar
mass by accretion from its companion (see Nomoto et al. 1997;
Hillebrandt \& Niemeyer 2000; Livio 2001 for reviews). However, the
progenitor binary system leading to SNe Ia is still unknown, and there
are two competing scenarios for the accretion process triggering SNe
Ia. In the single-degenerate (SD) scenario, the accretion is from a
non-degenerate companion star (Whelan \& Iben 1973; Nomoto 1982),
while in the double-degenerate (DD) scenario, a merger of two white
dwarfs results in a SN Ia (Iben \& Tutukov 1984; Webbink 1984).  To
reveal the progenitor is important not only for better understanding
of this one of the brightest explosions in the universe, but also for
controlling systematic uncertainties when SNe Ia are used as a
standard candle to measure the expansion rate of the universe (Riess
et al. 1998; Perlmutter et al. 1999). SNe Ia are expected to have a
wide range of delay time from star formation to supernova explosions,
and the delay time distribution (DTD) can be used to discriminate the
proposed progenitor models, since different progenitor scenarios
predict different DTDs.

The DTD $f_D(t_{\rm Ia})$ is equivalent to the SN Ia occurrence rate
as a function of the Ia delay time $t_{\rm Ia}$, for a single-burst
stellar population of a unit stellar mass. Hence, observational
studies on SN Ia rates should be able to constrain DTD (Mannucci et
al. 2006 and references therein). One such approach is to examine the
evolution of the cosmic SN Ia rate density (Pain et al. 2002; Gal-Yam
\& Maoz 2004; Maoz \& Gal-Yam 2004; Strolger et al. 2004; Dahlen et
al. 2004; Barris \& Tonry 2006; F\"orster et al. 2006; Neill et
al. 2006; Botticella et al. 2008; Oda et al. 2008; Blanc \& Greggio
2008).  However, there is a degeneracy between SN Ia DTD and the
cosmic star formation history, and it is difficult to derive a strong
constraint on DTD from the currently available data (F\"orster et
al. 2006; Botticella et al. 2008; Oda et al. 2008; Blanc \& Greggio
2008).  Another approach is to study the dependence of SN rate on the
host galaxy properties, such as colors or spectral energy distribution
(SED) (Mannucci et al.  2005; Scannapieco \& Bildsten 2005; Sullivan
et al. 2006; Aubourg et al. 2007).  This approach has already given
some useful constraints on DTD, indicating the existence of SN Ia
populations having both short ($\lesssim$ 0.1 Gyr) and long ($\gtrsim$
10 Gyr) delay times.  However, the functional shape of $f_D(t_{\rm
  Ia})$ has not yet quantitatively been constrained well. What has
been done in previous studies is to assume simple mathematical
functions for DTD or adopt theoretical DTD models to predict the
distributions of some observational quantities such as host galaxy
colors, and then compare them to the observed data. A clear next step
is a measurement of DTD, i.e., to constrain the functional form of DTD
directly from the observed data, rather than testing particular DTD
functions or models.

For a DTD measurement, one needs to reliably estimate the delay time
of an observed SN Ia and the stellar mass of its host galaxy. The mean
stellar age of the host galaxy is an indicator of the delay time, but
it is unreliable when the stellar age distribution has a large
dispersion, as in galaxies having extended star formation history.
The ideal population for this purpose is old galaxies with an
approximately uniform stellar age, which experienced major star
formation episode in the past and have little or no star formation
activity at the time of a SN explosion. Present-day elliptical
galaxies are good examples, but they have a typical age of $\gtrsim$
10 Gyr (Barber et al. 2007; Jimenez et al. 2007), and hence we need to
observe SNe at higher redshifts to get a sample of SNe having shorter
delay times. However, both detection of SNe and stellar age estimate
of host galaxies become difficult at high redshifts.

The Subaru/XMM-Newton Deep Survey (SXDS, Furusawa et al. 2008), which
is the deepest survey among those wider than $\sim$1 deg$^2$ with a
broad coverage of various wavelengths, provides a unique data set for
this purpose. In optical bands, this field has been observed
repeatedly with some time intervals, and a systematic variable object
search has been performed (Morokuma et al. 2008a, b), leading to
detection of more than 1,000 variable objects down to a limiting
variability magnitude of $m_{i'} \sim 25.5$ (AB). The majority of them
are high-redshift supernovae and active galactic nuclei (AGNs). This
data set includes many passively evolving old galaxies at redshift $z
\sim 1$, which are believed to be the direct ancestors of the
present-day elliptical galaxies (Yamada et al. 2005).  We can estimate
the stellar age and stellar mass of host galaxies by photometric
redshift calculations using the rich photometric data in the 9 band
filters ($BVR_c \ i'z'JK$, 3.6, and $4.5 \mu$m).  The required
accuracy for the age estimate in this work is about a factor of two,
and we will show that such an accuracy can be achieved for the
galaxies used in this work, by a variety of tests for the age
estimates.

In this paper we measure the SN Ia DTD by selecting the SXDS variable
objects found in old or passively evolving galaxies at $z \sim $
0.4--1.2. Spectroscopic SN type confirmation is not available for the
faint SXDS SN candidates. It is difficult to construct a complete
high-redshift SN sample with spectroscopic type classification under
homogeneous conditions of spectroscopic observation, and contamination
of SNe with no or poor spectroscopic data is one of the most
challenging sources of uncertainty in SN rate studies (Strolger et
al. 2004; Neill et al. 2006; Sullivan et al. 2006; Poznanski et
al. 2007).  Here, instead of using spectroscopic information, we
select variable objects in old galaxies with spatial offset from the
galactic centers, and hence the majority of them are expected to be
SNe Ia. In fact, we will quantitatively demonstrate that more than
80 \% of them should be SNe Ia, based on the properties of
the SN candidates and their host galaxies.

It should be noted that selecting SNe only in a particular type of
galaxies does not induce bias in the DTD estimates, since we measure
SN Ia rate normalized by stellar mass of host galaxies having the same
type. This is in contrast to measurements of total cosmic SN rate
density, in which selection of any particular galaxy type obviously
leads to an underestimate of the total rate. On the other hand, we
cannot measure DTD at short delay times obviously because of selecting
old galaxies. We will present a measurement of DTD in a range of
$t_{\rm Ia} $ = 0.1--8.0 Gyr, corresponding to the distribution of
mean stellar ages of the old galaxies used in this work. Our DTD
measurement will be supplemented by the SN Ia rate measured for
elliptical galaxies in the local universe, to obtain DTD in $t_{\rm Ia}
\sim $ 0.1--10 Gyr. The derived DTD will then be compared with a wide
range of the existing theoretical DTD predictions, to get implications
for the SN Ia progenitor.

In section \ref{section:methods}, we describe the SXDS data set and
selection procedures of old galaxies and SN candidates.  In section
\ref{section:DTD}, we describe the formulations of DTD measurement and
present the results.  In section \ref{section:systematics}, we examine
various systematic uncertainties in our DTD estimates.  We then
discuss about the implications for the SN Ia progenitor, from the
comparison between the measured DTD and theoretical predictions
(section \ref{section:models}).  Summary and conclusions are given in
section \ref{section:conclusions} with some discussions.  Throughout
this paper we use the standard $\Lambda$CDM cosmological parameters of
$(h, \Omega_M, \Omega_\Lambda) = (0.7, 0.27, 0.73)$, where $h \equiv
H_0$/(100 km/s/Mpc), and magnitudes are given in the AB magnitude
system unless otherwise stated.

\section{The SXDS SN Ia Candidates}
\label{section:methods}

\begin{figure*}
\begin{center}
\includegraphics[width=150mm,angle=-90]{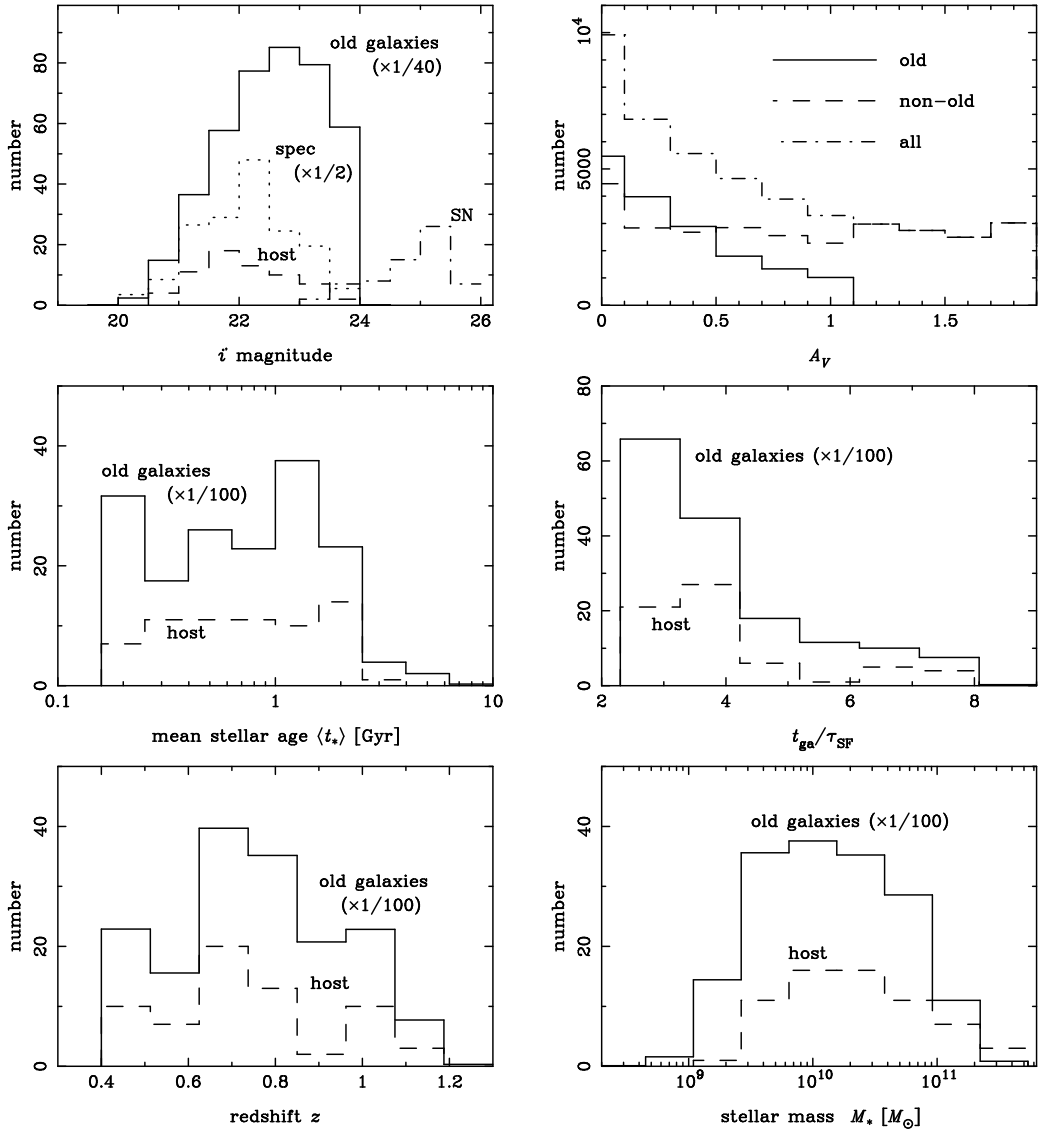} 
\end{center}
\caption{The properties of the old galaxies selected for the SN
  search.  Upper left panel: the $i'$-band magnitude distributions of
  all the 16,492 old galaxies used in the analysis (solid), the 314
  old galaxies with spectroscopic data (dotted), and the host galaxies
  of the 65 SN candidates (dashed). The dot-dashed line is the
  distribution of the $i'$-band variability magnitude of the 65 SN
  candidates. Upper right panel: the $A_V$ distributions of the old
  galaxies (solid), non-old galaxies (dashed), and all (old plus
  non-old) galaxies (dot-dashed) with $m_{i'} \leq 24.0$ and
  $m_{3.6\mu \rm m} \leq 22.8$. Middle and lower panels: the
  distributions of $\langle t_* \rangle$, $t_{\rm ga}/\tau_{\rm SF}$,
  redshift, and stellar mass of all the old galaxies (solid) and the
  host galaxies of the SN candidates (dashed).}  \label{fig:i_hist}
\end{figure*}

\subsection{the SXDS Data and Photometric Redshift Calculations}
\label{section:photo-z}

We utilize the rich photometric data set of SXDS in $BVR_c \ i'z'$
bands obtained by the Subaru/Suprime-Cam (Furusawa et al. 2008), in
$JK$ bands obtained by the UKIDSS survey (Warren et al. 2007), and in
3.6 and 4.5 $\mu$m bands obtained by the SWIRE survey (Lonsdale et
al. 2004), with the limiting magnitudes ($3 \sigma$) of 28.4, 27.8,
27.7, 27.7, 26.6, 24.1, 24.0, 23.1, and 22.4, respectively, for
aperture diameters of 2 (optical and $JK$) and 3.8 (the Spitzer bands)
arcsec. The point-spread-function size of the optical images is about
0.8 arcsec FWHM (1 pixel = 0.2 arcsec).  The total survey area used in
the variable object survey (Morokuma et al. 2008a) is 0.918 deg$^2$.
Though the $JK$ data is available only for 64.5\% of the SXDS
field, almost all (97.5\%) of the SXDS field is covered by the Spitzer
data. This is important for reliable stellar mass estimate, which is
crucial for a study of SN Ia DTD. Therefore we select relatively
bright 45,374 galaxies with $m_{i'} \leq 24.0$ and $m_{\rm 3.6\mu m}
\leq 22.8$, for reliable photo-$z$ and stellar mass/age estimations.
Stars have been removed by the two-color plot ($m_{R_c}-m_{i'}$ versus
$m_{R_c}-m_{3.6\mu \rm m}$) as in Morokuma et al. (2008a).

Photometric redshift ($z_{\rm ph}$) calculations are performed by
using the publicly available code {\it hyperz} (Bolzonella, Miralles,
\& Pell 2000), with the GALAXEV library (Bruzual \& Charlot 2003) of
stellar population synthesis models based on the Padova 1994
evolutionary tracks.  In our baseline analysis, we use the library
assuming the Salpeter initial mass function (IMF) in the mass range of
0.1--100 $M_\odot$ and a supersolar abundance of $Z = 0.05$, since
massive galaxies with old stellar population are expected to have high
metallicity as known for the local elliptical galaxies (Barber et
al. 2007; Jimenez et al. 2007). We use 7 models of star formation (SF)
history having exponentially decaying star formation rate (SFR) with
the exponential time scales of $\tau_{\rm SF} = $ 0.1, 0.3, 0.7, 1, 3,
5, and 15 Gyrs.  We also add a model with a constant SFR, and hence 8
SF history templates are used in total. Attenuation of galaxy spectra
by dust is taken into account with the Calzetti law (Calzetti et
al. 2000) within a range of $A_V = $ 0--2, where $A_V$ is the
restframe visual band attenuation.  The minimum photometric error is
set to be $\pm 0.05$ mag, to include systematic errors for bright
objects having negligibly small statistical photometric errors.  Then the
best-fit redshift, SF history template, galaxy age, and $A_V$ are
calculated for each galaxy. The age survey range is limited not to
exceed the age of the universe at a given redshift.

\subsection{Old Galaxy Selection}
\label{section:old_galaxies}

We select old or passively evolving galaxies by requiring $t_{\rm ga}
/ \tau_{\rm SF} > 2.3$, where $t_{\rm ga}$ is the galaxy age estimated
by the {\it hyperz} code, i.e., the time elapsed from the beginning of
the template SF history. This value is chosen so that
galaxies selected by this criterion have already formed more than $[1
- \exp(-2.3)] = $ 90\% of stars that the galaxies would form in all
the history.  We also set a constraint of $A_V \leq 1.0$ to avoid
dusty objects.  After selecting galaxies within $0.4 \leq z_{\rm ph}
\leq 1.2$, which is a typical redshift range expected for SNe Ia
detectable in SXDS, there remain 16,492 galaxies. We call these
galaxies simply ``the old galaxies'' in this work for convenience.

Stellar masses are calculated for each of the old galaxies from the
results of the photo-$z$ fit.  In this work, we define the stellar
mass of a galaxy, $M_*$, as the integration of SFR up to its
age. Because dying stars return a part of their mass into interstellar
medium, this is not exactly the same as the mass locked up in stars at
a given age, which changes with time even in a passive evolution
phase.  [The difference between the two is typically $\lesssim$ 30\%;
see, e.g., Fig. 7 of Sullivan et al. (2006).] It should be noted that
the SFR-integrated mass is more appropriate for our purpose, because
DTD should be normalized by the amount of star formation for a single
starburst stellar population, which is constant against age. However,
stellar mass estimations for observed galaxies are generally
model-dependent.  Therefore, when we derive DTD, we convert the
stellar mass of our definition into $L_{K,0}$, which is the restframe
$K$ band luminosity at the age of 11 Gyr (a typical age of local
elliptical galaxies; see e.g., Baber et al. 2007; Jimenez et al. 2007)
after passive evolution, since restframe $K$ band luminosity at a
fixed large age is a good indicator of stellar mass.  Then DTD will be
given per unit $L_{K,0}$, and hence the difference in mass definition
does not affect our DTD measurements. By this normalization, we can
avoid some observational uncertainties in stellar mass estimates such
as IMF (see also section \ref{section:syst_age_mass}), and a
comparison between our results and other SN rate studies also becomes
easier.

\begin{figure}
\begin{center}
\includegraphics[width=6cm,angle=-90]{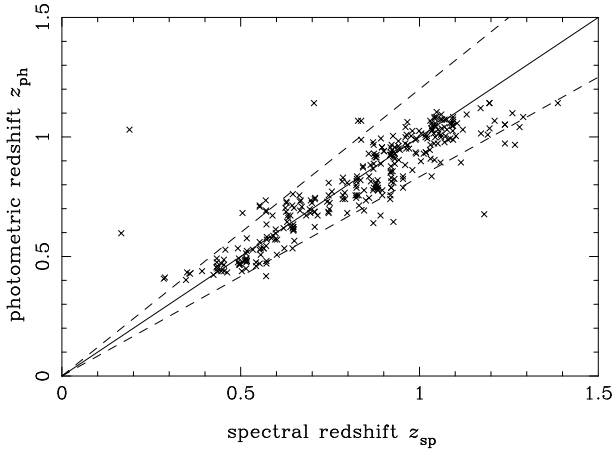} 
\end{center}
\caption{ Spectral redshift versus photometric redshift of the old
galaxies selected for the SN Ia search. The dashed lines indicate the
$\pm$20\% accuracy region.}  \label{fig:z_sp_z_ph}
\end{figure}

The distributions of $i'$ magnitude, $A_V$, mass-weighted mean stellar
age $\langle t_* \rangle$, $t_{\rm ga}/\tau_{\rm SF}$, redshift, and
stellar mass of these old galaxies are shown in Fig. \ref{fig:i_hist}.
As expected for old galaxies, the distribution of $A_V$ is peaked at
$A_V=0$, with a mean value of $\langle A_V \rangle = $ 0.31. The
concentration to small $A_V$ values is obvious when it is compared
with the distribution of the non-old galaxies.  To test the
reliability of photo-$z$, we compare the results with the
spectroscopic redshifts ($z_{\rm sp}$) of the 314 old galaxies having
observed spectra in Fig. \ref{fig:z_sp_z_ph}. The agreement is within
$\pm$20\% for the majority of the old galaxies. As a measure of
deviation that is not sensitive to outliers, we computed the median of
$1.48 |\Delta z|/(1+z)$, which is the same as the standard deviation
of $\Delta z /(1+z)$ in the case of the Gaussian distribution.  The
result is 0.035, and this is comparable with other photometric
redshift studies at similar redshifts (e.g., Ilbert et al. 2006).  The
magnitude distribution of the spectroscopic sample peaks at $m_{i'}
\sim 22$--23 and extends down to $m_{i'} \sim 24$
(Fig. \ref{fig:i_hist}), and hence our photometric old galaxy sample
is restricted to the magnitude range where the spectroscopic
calibration is possible.  In the DTD analysis below, we use
spectroscopic redshifts when available, and otherwise use photometric
redshifts.

\begin{figure*}
\begin{center}
\includegraphics[width=150mm,angle=0]{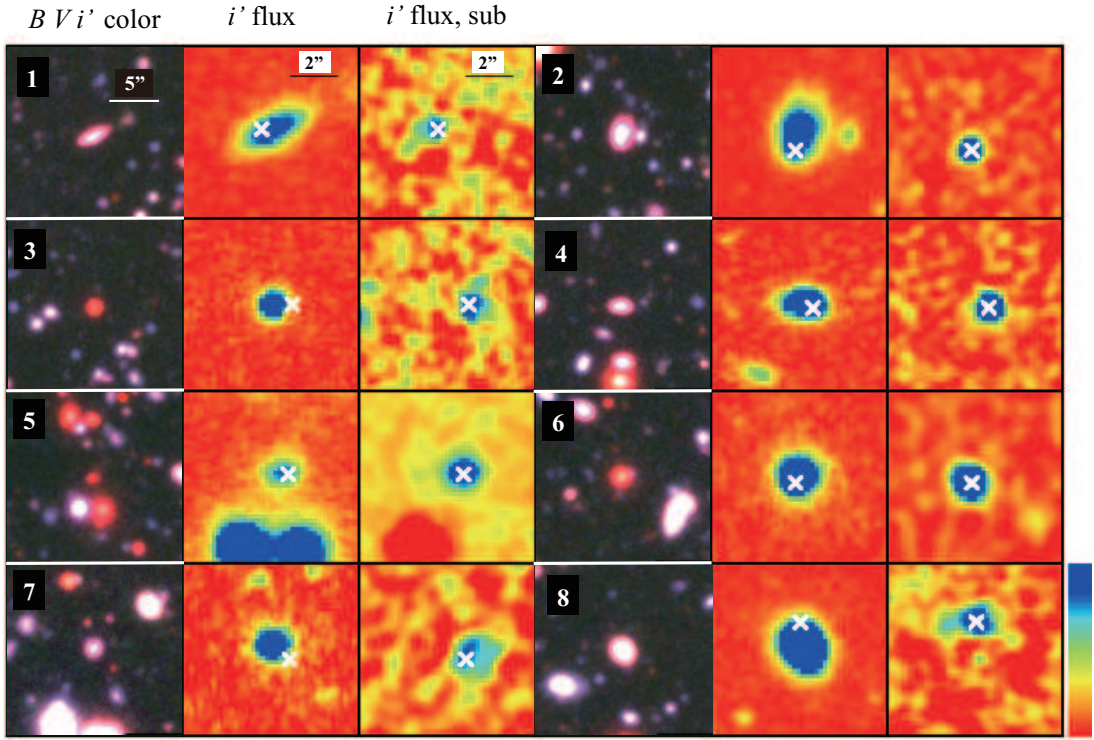}
\end{center}
\caption{The images of the eight examples randomly selected from the
final SN candidates. The left panels are the 3-color composite images of
the host galaxies using $BVi'$ bands.  The middle panels are color
contours of the host galaxy $i'$ band flux at the time of the maximum
luminosity of the SN candidates, and the right panels are those of
subtracted images showing the variability of the SN candidates. All
panels are centered at the surface brightness peak of the host galaxies,
and the locations of the SN candidates are indicated by white crosses.
The properties of the eight objects are summarized in Table
\ref{table:examples}. } \label{fig:images}
\end{figure*}

\begin{figure*}
\begin{center}
\includegraphics[width=150mm,angle=-90]{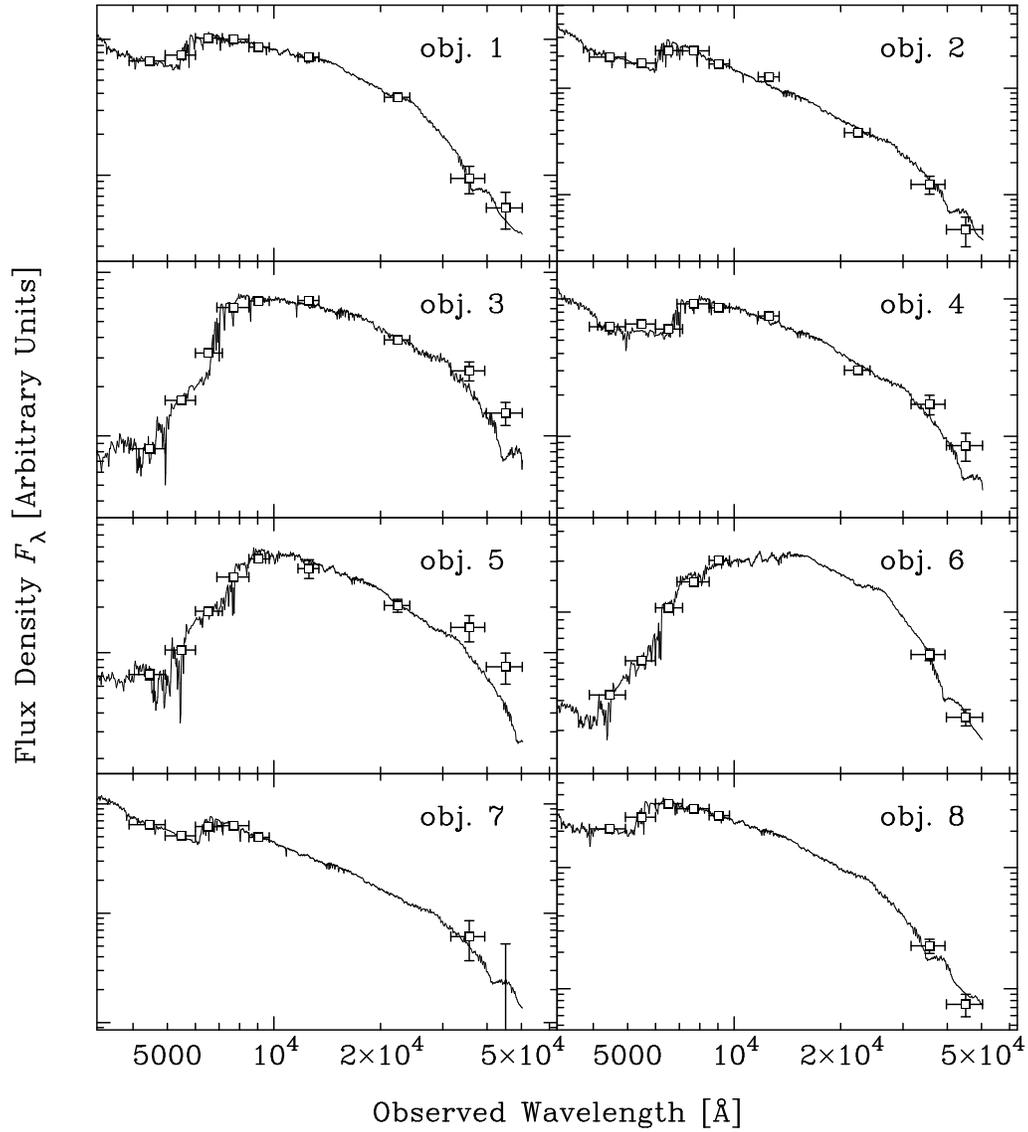} 
\end{center}
\caption{The results of the SED fits for the old galaxies by the
  photometric redshift calculations, for the 8 example objects listed
  in Table \ref{table:examples}. The SED is given by energy flux per
  unit wavelength in arbitrary units.  The open squares are the
  observed flux. The horizontal error bars indicate the approximate
  width of band filters, and the vertical error bars are 1$\sigma$
  flux errors (difficult to see in optical bands).  The flux errors
  include only statistical errors, while the minimum error is set to
  be $\pm 0.05$ mag in the photometric redshift calculations to take
  into account the systematic errors.  }
\label{fig:SED}
\end{figure*}

For visual demonstrations, we randomly selected 8 objects as examples
from our final sample of SN candidates. The properties of the 8
objects are summarized in Table \ref{table:examples}. The host galaxy
images and the SED fits of the 8 objects are shown in
Figs. \ref{fig:images} and \ref{fig:SED}, respectively.  The
photometric errors are very small especially in the optical bands
since we selected bright objects, but still the agreement between the
template SEDs and the observed data is quite good, which is an
encouraging result about the reliability of age and stellar mass
estimates.

\subsection{Selection of the SN Candidates}

We now search for variable objects associated with these old galaxies.
There are 1,040 variable objects selected by $i'$ band variability in
the catalog of Morokuma et al. (2008a). The SXDS field was monitored
8--10 times during about four years, with a typical time interval from
a few days to one month within a year. The variable sources are
selected if it is detected in the subtracted images of any possible
pair of two different epochs.  The detection efficiency of variable
sources has been carefully determined by simulations.  ``The
variability flux'' of a SN candidate in this work refers to the
maximum of differential fluxes measured in time intervals
corresponding to all possible pairs of the SXDS observing epochs.
``The variability magnitude'' is the magnitude corresponding to this
variability flux.

We select variable objects within the detection isophote (i.e., the
area where the surface brightness is higher than the threshold level
of the source detection) of the old galaxies, and showing significant
offsets ($\geq$ 1.9 pixel = $0.38''$) from the nuclei of galaxies.
The mean area of detection isophote of the old galaxies is 409 pixels.
Here, the nuclei of galaxies are simply defined by their surface
brightness peaks. According to the simulations using artificial
objects (Morokuma et al. 2008a), this is a conservative limit to
remove variable objects at nuclei of the host galaxies, i.e., AGNs.
Morokuma et al. classified the nuclear variable sources into SNe or
AGNs based on the light-curve shape, and they found that such
classifications are broadly consistent with the X-ray
information. However, the classification is rather uncertain
especially for the faintest variable objects near the detection
limit. Therefore we conservatively reject all nuclear variable
sources. We do not use the X-ray information in the process of SN
candidate selection, since a small part ($\sim$ 10\%) of the survey
area is not covered by the X-ray observation. Galaxies detected in the
X-ray band are only 68 out of the 14,909 old galaxies observed in
X-ray, which is a negligible fraction in our analysis. In fact, we
confirmed that no object is detected in X-ray in our final SN
candidate sample.

We thus found 67 variable objects associated with the old galaxies.
We examined the optical variability information of these objects by
using the full Suprime-Cam data set of the SXDS. We found that, as
expected, most of them are consistent with SN-like variability. Only 2
objects show clearly AGN-like variability, i.e., flaring up more than
twice during the four year observation duration. These two are
removed, leaving the 65 SN candidates. Such a contamination of AGNs is
possible by a chance superposition of a normal galaxy and an unrelated
AGN along the line of sight (e.g., the case of an AGN initially
classified as SN 1999de, Gal-Yam et al. 2008).  We found that two such
events are reasonable from the number of AGNs in SXDS and the area
covered by the old galaxies selected here (2.3\% of the total survey
area). We cannot exclude a possibility that a few more
chance-superposition AGNs having SN-like light curves might be
included in the 65 SN candidates. To check this, we examine the offset
distribution of these objects with respect to the surface brightness
profile of galaxies as follows.

\begin{table*}
  \caption{Properties of the eight examples randomly selected from the
    final 65 SN sample. The redshifts are spectroscopic when available
    (objects 2 and 4), and otherwise photometric.  Magnitudes of host
    galaxies and SN variability are in $i'$ band, and galaxy age ($t_{\rm
      ga}$), the exponential decay time scale of SFR ($\tau_{\rm SF}$), and
    mass-weighted mean stellar age ($\langle t_* \rangle$) are all in Gyrs.}
\begin{center}
\begin{tabular}{cccccccc}
\hline
\hline
Obj. No. & redshift & \ host mag. \ & \ var. mag. \ & \ $t_{\rm ga}$ \
& \ $\tau_{\rm SF}$ \ & \ $\langle t_* \rangle$ \ & \ $A_V$ \ \\
\hline
1 & 0.47 & 22.5 & 25.4 & 2.30 & 1.0 & 1.56 & 0.2 \\
2 & 0.63 & 21.6 & 24.6 & 0.36 & 0.1 & 0.27 & 0.0 \\
3 & 0.81 & 23.0 & 25.8 & 0.72 & 0.1 & 0.62 & 0.8 \\
4 & 0.92 & 22.6 & 24.2 & 1.02 & 0.3 & 0.75 & 0.0 \\
5 & 1.01 & 23.7 & 24.5 & 0.72 & 0.1 & 0.62 & 0.2 \\
6 & 0.60 & 22.0 & 23.5 & 3.50 & 0.7 & 2.82 & 0.6 \\
7 & 0.70 & 23.0 & 25.2 & 0.72 & 0.3 & 0.49 & 0.0 \\
8 & 0.47 & 21.3 & 25.1 & 1.02 & 0.3 & 0.75 & 0.0 \\
\hline
\hline
\end{tabular}
\end{center}
\label{table:examples}
\end{table*}

\subsection{Radial Distribution of the SN Candidates in Host Galaxies}

\begin{figure*}
\begin{center}
\includegraphics[width=60mm,angle=-90]{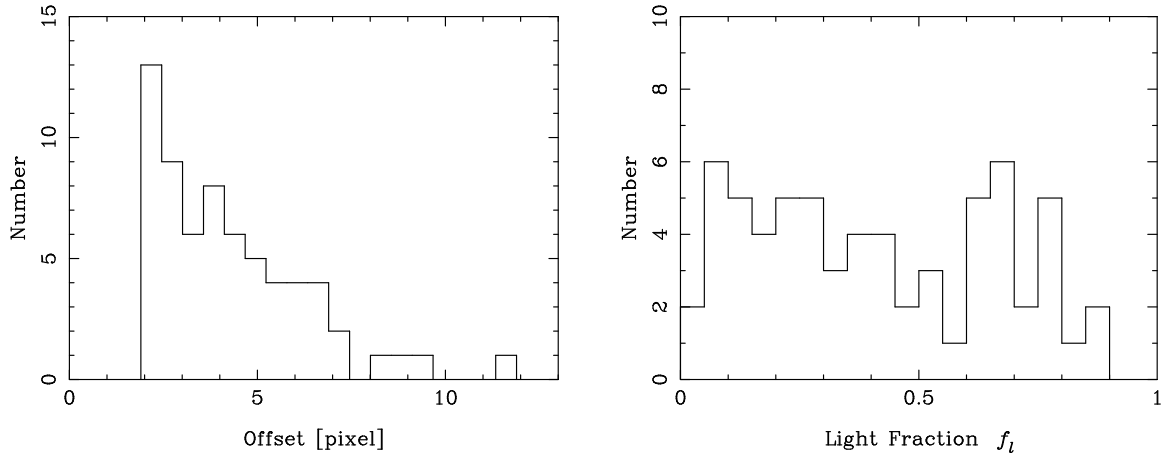} 
\end{center}
\caption{Left panel: the distribution of the spatial offsets of
the 65 SN candidates from the nuclei of their host galaxies.  Right
panel: the distribution of the fraction of $i'$-band
galactic light $f_l$ enclosed
within the ellipsoidal radius to a SN candidate compared with the
isophotal flux of its host galaxy.}
\label{fig:profiles}
\end{figure*}

We performed ellipsoidal fits for each galaxy with the S\'ersic type
radial profile (S\'ersic 1968).  The effective radius along the major
and minor axes, and the S\'ersic index are thus derived, taking into
account the seeing. We then calculated the fraction, $f_l$, of
$i'$-band light enclosed within the ellipsoidal radius to a variable
object, compared with the total flux within the detection isophote
(i.e., isophotal flux).  Here, the nuclear region within the offset
threshold (1.9 pixel radius) is excluded in the $f_l$ calculation. If
we are selecting SNe Ia physically associated with the galaxies, we
expect that the distribution of $f_l$ is roughly uniform between 0--1, since
SNe Ia are expected to trace the host galactic light approximately 
(Kelly et al. 2007; F\"orster \& Schawinski 2008).  The distributions
of the offsets from nuclei and $f_l$ are shown for the 65 SN
candidates in Fig. \ref{fig:profiles}, and in fact we find an almost
uniform distribution for $f_l$.

There is a deficit at $f_l \gtrsim 0.8$, and it indicates that our
profile fitting is not perfect or supernova locations do not exactly
trace galactic $i'$-band light. However, it is clear that the SN
candidates trace the galactic light reasonably well, and hence the
majority of these must be SNe physically associated with host
galaxies, rather than the chance superpositions of background AGNs. It
should be noted that, if there is a significant contamination of AGN
chance superpositions, we expect a distribution biased to larger
values of $f_l$, which is opposite from the observed trend. We thus
define these 65 objects as the final SN candidate sample, and the
distributions of variability and host galaxy magnitudes are shown in
Fig.  \ref{fig:i_hist}. The images of the randomly selected eight
examples are shown in Fig. \ref{fig:images}.

In the final SN sample, 7 SN candidates have measured spectroscopic
redshifts of host galaxies.  We confirmed that the difference between
$z_{\rm sp}$ and $z_{\rm ph}$ is within $\sim$20 \% for all of these, as
expected from Fig. \ref{fig:z_sp_z_ph}.  The fraction of available
spectroscopic redshifts, 7/65, is considerably higher than 314/16492 for
the old galaxies. This is mainly because the host galaxies of the SN
candidates are systematically brighter than the typical old galaxies
(see Fig. \ref{fig:i_hist}), reflecting the fact that SN rate is roughly
proportional to host galaxy luminosities. Another possible reason is
that some variable objects were followed up by spectroscopic
observations as a part of the Supernova Cosmology Project in
collaboration with the SXDS project.

\section{Measurement of Delay Time Distribution}
\label{section:DTD}

\subsection{The Basic Formulations}
\label{section:formulations} 

Now we estimate the DTD from the 65 SN candidates.  Later (section
\ref{section:systematics}) we will demonstrate that the majority of
them must be SNe Ia rather than core-collapse (CC) SNe, but here we
first estimate the DTD, taking into account the contamination of CC
SNe.  We implicitly assume that DTD is universal for all stellar
populations, as often assumed in studies on the SN Ia DTD.  It is
possible that DTD is dependent on the properties of stellar population
such as metallicity (e.g., Kobayashi et al. 1998), and in such a case
our measurement of DTD applies only to the old galaxies selected
here. This point should be kept in mind when a comparison is made with
theoretical predictions.

Since we know the star formation history of the template galaxy
evolution model in the {\it hyperz} fit, we can estimate the
mass-weighted mean stellar age of galaxies, $\langle t_* \rangle$.
Since we selected old galaxies, in which most of stars have formed in
a starburst in the past, $\langle t_* \rangle$ is expected to be a
good estimator of the Ia delay time, $t_{\rm Ia}$.  (By the
same reason, the difference between the mass-weighted and
luminosity-weighted mean stellar ages should be small.)  The DTD
function $f_D (t_{\rm Ia})$ (per unit stellar mass and unit
delay time) can be estimated as the SN Ia rate per unit stellar mass
in these galaxies.  Therefore the first rough estimate of the mean DTD
function $\bar{f}_D$ in a given bin of $t_l \leq t_{\rm Ia} < t_u$ can
be obtained by solving the following equation:
\begin{eqnarray}
N_{\rm Ia, exp} + N_{\rm CC, exp} = N_{\rm obs} \ , 
\label{eq:N_SN}
\end{eqnarray}
where $N_{\rm Ia, exp}$ and $N_{\rm CC, exp}$ are the expected numbers
of Ia and CC SNe, respectively, and $N_{\rm obs}$ is the number of the
observed SN candidates associated with the old galaxies satisfying $t_l
\leq \langle t_* \rangle < t_u$. We can calculate $N_{\rm Ia, exp}$ as:
\begin{eqnarray}
N_{\rm Ia, exp} = 
\bar{f}_D \sum_{i} f_{\rm off, \it i} \ M_{*, i} \ 
\frac{T_{V, {\rm Ia}}(z_i)}{(1 + z_i)} \
\label{eq:N_Ia_exp} \ ,
\end{eqnarray}
where $M_{*, i}$ and $z_i$ are the stellar mass and redshift of $i$-th
galaxy, $f_{\rm off, \it i}$ the fraction of light (in $i'$ band)
outside the threshold offset (1.9 pixel radius), and the summation is
for all the old galaxies satisfying $t_l \leq \langle t_* \rangle <
t_u$. In this work, we set the shortest delay time bin to be 0.1--0.25
Gyr.  This is reasonable because the minimum value of $\langle t_*
\rangle$ possible for the old galaxies selected by our criteria is
0.16 Gyr corresponding to the case of $t_{\rm ga} = $ 0.23 Gyr and
$\tau_{\rm SF}$ = 0.1 Gyr, with an age dispersion of about 0.1 Gyr.
On the other hand, the number of the old galaxies with $\langle t_*
\rangle > $ 8.0 Gyr is very small because they are located at $z \ge
0.4$, and hence we set the upper boundary of our DTD measurement to be
8.0 Gyr.

\begin{figure}
\begin{center}
\includegraphics[width=60mm,angle=-90]{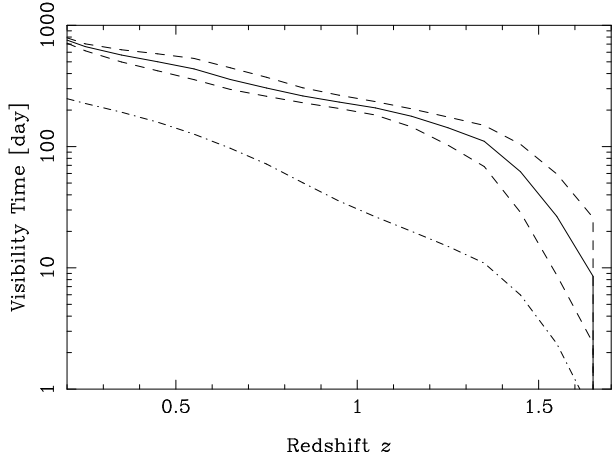} 
\end{center}
\caption{The visibility time $T_V$ (in observer's frame) of SNe in SXDS
as a function of redshift. The solid curve is for the standard case of
SNe Ia, while the dot-dashed curve is for CC SNe. 
The dashed curves are for SNe Ia, but the luminosity is
shifted by $\pm$ 0.4 mag.  The visibility time is slightly different in
the different subfields of SXDS, and those in the SXDS-C field (Morokuma
et al. 2008a) are shown here. } \label{fig:vt}
\end{figure}

The visibility time $T_V$ (also often called as the control time),
which is the total integrated time during which a SN can be detected
by the SXDS observations, can be calculated as:
\begin{eqnarray}
T_V(z) = \int_{-\infty}^\infty  P_{\rm det}(t, z) dt \ ,
\label{eq:T_V}
\end{eqnarray}
where $t$ is the explosion time of a supernova relative to the SXDS
observation campaign (in observer's frame), and $P_{\rm det} (t, z)$
is the detection probability of the supernova at redshift $z$. The
factor $(1+z_i)^{-1}$ in eq. (\ref{eq:N_Ia_exp}) corrects the
cosmological time dilation effect. This is the standard quantity in SN
rate analyses, and it can be calculated if the SN light curve is given
in the observing band filter.  We apply the standard light curve and
color evolution of SNe Ia used in Oda \& Totani (2005), taking into
account the dispersion of the peak $B$ band luminosity and the
light-curve versus peak luminosity relation. We assume an extinction
corresponding to the reddening $E(B-V) = 0.05$, which is reasonable
for old galaxies and often used in rate studies of SNe Ia [see Oda et
al. (2008) and references therein]. We assume the standard Milky-Way
extinction curve. It should be noted that the Calzetti {\it
  attenuation} law used in the photo-$z$ calculations is for the
synthesized flux from a galaxy, which is different from the {\it
  extinction} law for a single star in a galaxy. Therefore $A_V$ of a
galaxy estimated by photo-$z$ calculation cannot be directly related
to extinction of a SN in the galaxy. The effect of changing extinction
of SN flux will be examined in section \ref{section:systematics}.
Finally, the exact procedure and efficiency of variable object
detection in SXDS (Morokuma et al. 2008a) are also taken into account
to calculate $P_{\rm det}$.  The calculated visibility time is shown
in Fig. \ref{fig:vt}. Note that this visibility time includes the
multiplicity of many SXDS observations during four years, and not for
a single epoch observation.

The expected number of CC SNe,  $N_{\rm CC, exp}$, can be calculated as
\begin{eqnarray}
N_{\rm CC, exp} = \sum_i \ f_{\rm off, \it i}  \ \psi_i \ f_{\rm CC} 
\ \frac{T_{V, {\rm CC}}(z_i)}{(1+z_i)} \ ,
\end{eqnarray}
where $\psi$ is SFR (mass per unit time) of a galaxy, and $f_{\rm CC}$
the production efficiency of CC SNe per unit mass of star formation.
The SFR $\psi$ is estimated by the observed $B$ band flux corresponding
to the restframe UV luminosity, with the conversion factor calculated by
the stellar population synthesis model used in this work.  The factor
$f_{\rm CC}$ is calculated by assuming that all stars heavier than 8
$M_\odot$ produce CC SNe with the assumed IMF. The visibility time
$T_{V, \rm CC}$ is calculated with a standard mixture of light curves of
various types of CC SNe, as in Oda \& Totani (2005). We assume $E(B-V) =
0.15$ for CC SNe, which is a typical value for CC SNe found in nearby
galaxies (Oda et al. 2008 and references therein). This value is larger
than that assumed for SNe Ia, but it is reasonable since CC SNe are
expected to occur in star forming regions where dust abundance is
generally high. Since we selected old galaxies, we found $N_{\rm CC,
exp} = 11.4$ (in all the considered range of $t_{\rm Ia}$ = 0.1--8 Gyr) and
this is small compared with $N_{\rm obs} = 65$. (See also Table
\ref{table:baseline} for numbers in each $t_{\rm Ia}$ bin.)

\begin{table*}
  \caption{The DTD measurements by the baseline analysis.  The DTD results
    are shown in [century$^{-1}$] for a single starburst population whose
    total $K$ band luminosity is $10^{10} L_{K, \odot}$ at an age of 11
    Gyr. The errors are statistical $1 \sigma$. The number of the old galaxies
    ($N_{\rm gal}$), that of the detected SN candidates ($N_{\rm obs}$), and
    the expected numbers of the prompt SNe Ia ($N_{\rm pIa, exp}$) and CC
    SNe ($N_{\rm CC, exp}$) are shown. The last column shows the mean of
    $\sigma_{t_{\rm Ia}}$ of galaxies in a time bin, where $\sigma_{t_{\rm
        Ia}}$ is the standard deviation of the probability distribution of 
    $t_{\rm Ia}$ in a galaxy.}
\begin{center}
\begin{tabular}{ccccccc}
\hline
\hline
$t_{\rm Ia}$ bin [Gyr] & DTD $f_D(t_{\rm Ia})$ \ \ & $N_{\rm gal}$ \ \ &
$N_{\rm obs}$ \ \ & $N_{\rm pIa, exp} \ \ $ 
& $N_{\rm CC, exp}$ \ \ & $\langle \sigma_{t_{\rm Ia}} \rangle $ [Gyr] \\
\hline
0.1--0.25 &  $2.89^{+1.60}_{-1.20}$ &  3719 & 12 & 1.0 & 2.7 & 0.05 \\
0.25--0.5 & $1.98^{+0.96}_{-0.73}$ &  3086 & 14 & 0.92 & 3.1 & 0.14 \\
0.5--1.0 & $0.84^{+0.30}_{-0.24}$ & 4260 & 19 & 0.88 & 3.0 & 0.30 \\
1.0--2.0 & $0.49^{+0.17}_{-0.13}$ & 4764 & 19 & 0.66 & 2.3 & 0.50 \\
2.0--4.0 & $0.12^{+0.38}_{-0.12}$ & 564 & 1 & 0.051 & 0.20 & 1.2 \\
4.0--8.0 & $0.00^{+1.08}_{-0.00}$ & 99  & 0 & 0.007 & 0.032 & 2.1 \\
\hline
\hline
\end{tabular}
\end{center}
\label{table:baseline}
\end{table*}

\begin{figure}
\begin{center}
\includegraphics[width=60mm,angle=-90]{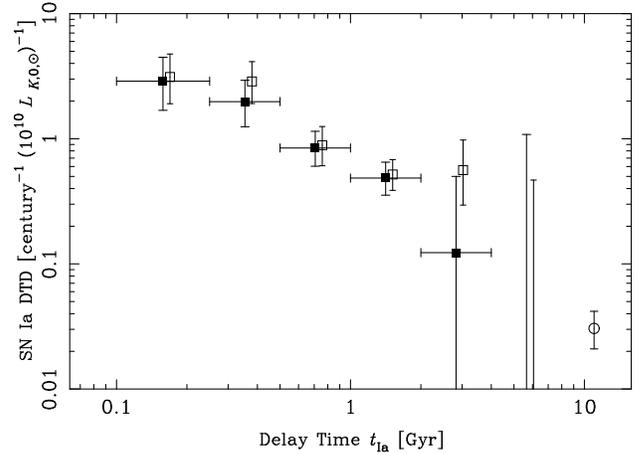}
\end{center}
\caption{The SNe Ia DTD, $f_D(t_{\rm Ia})$, per unit delay
  time $t_{\rm Ia}$ [century$^{-1}$] for a single starburst
  population whose total $K$-band luminosity is $10^{10} L_{K, \odot}$
  at the age of 11 Gyr.  The filled squares are the final
  observational estimates by this work based on the baseline analysis,
  and the error bars are statistical 1$\sigma$ errors. The open
  squares are the same but using a simpler method to estimate the
  delay time.  The time bins are the same as those for the filled
  squares, but the open squares are shifted in time by +0.03 dex in
  this plot for presentation.  The open circle is DTD inferred from
  the SN Ia rate in elliptical galaxies in the local universe
  (Mannucci et al. 2005). }
\label{fig:dtd_obs}
\end{figure}

The DTD estimated by this simple formulation is shown by open squares
in Fig. \ref{fig:dtd_obs}.  The error bars are statistical $1\sigma$
errors, which are calculated by the confidence limits of the small
number Poisson statistics (Gehrels 1986).  As mentioned in section
\ref{section:old_galaxies}, the normalization by stellar mass has been
converted into that by the $K$-band luminosity $L_{K,0}$, i.e., the
luminosity at the age of 11 Gyr after passive evolution. The
conversion is performed by calculating $M_*/L_{K,0}$ from the SED
templates used in the photo-$z$ calculations. For the template of
$\tau_{\rm SF} = $ 0.1 Gyr, we find $M_*/L_{K,0} = 1.8 \
[M_\odot/L_{K,\odot}$].  Here, we used $(V-K)_\odot = +1.49$ (Cox
2000) and $f_\lambda = 4.17 \times 10^{-11} \ \rm erg \ cm^{-2} s^{-1}
\AA^{-1}$ for $K=0$ (Zombeck 2007) in the Vega magnitude system.

We assume that the stellar age of nearby elliptical galaxies is 11
Gyr, based on the estimates for the SDSS early type galaxies (Barber
et al. 2007; Jimenez et al. 2007, and see section
\ref{section:local_data} for more details). Hence, the observed SN Ia
rate per unit $K$ luminosity in nearby elliptical galaxies,
$0.035^{+0.013}_{-0.011} (h/0.75)^2 \ {\rm century^{-1}} \
(10^{10}L_{K, \odot})^{-1}$ (Mannucci et al. 2005), gives an estimate
of $f_D(11 \rm \ Gyr)$, which can directly be compared with our
estimates. This value is based on the rate measurement of Cappellaro
et al. (1999) assuming $H_0 = $ 75 km/s/Mpc, and it is translated into
a value for $h=0.7$ used in this work.

Though these DTD results are derived by simply assuming $t_{\rm Ia} =
\langle t_* \rangle$, they are not significantly different from our
final results shown by the filled squares (see below), and well
described by a power-law, $f_D(t_{\rm Ia}) \propto {t_{\rm
Ia}}^{\alpha}$ with $\alpha \sim -1$ in 0.1--11 Gyr. However, to get a
more accurate estimate, we make a correction as follows.

\subsection{Correction for the Delay Time Estimates}

In the above formulation, we simply estimated $t_{\rm Ia}$ of an
observed SN Ia by $\langle t_* \rangle$ of its host galaxy. However,
in reality, there is a probability distribution of $t_{\rm Ia}$, which
is determined by the DTD and star formation history $\psi(t)$, where
$t = t_{\rm ga} - t_{\rm Ia}$ is the time elapsed from the beginning
of star formation in the host galaxy.  The expectation value of the Ia
delay time, $\langle t_{\rm Ia} \rangle$, is exactly the same as
$\langle t_* \rangle$ only when $f_D(t_{\rm Ia})$ is constant against
$t_{\rm Ia}$. The expectation value is generally given by:
\begin{eqnarray}
\langle t_{\rm Ia} \rangle = \frac{\int_0^{t_{\rm ga}} 
  \ t_{\rm Ia}
  \ \psi(t_{\rm ga} - t_{\rm Ia}) 
  \ f_D(t_{\rm Ia}) \ dt_{\rm Ia}  }{ \int_0^{t_{\rm ga}} 
  \ \psi(t_{\rm ga} - t_{\rm Ia}) \ f_D(t_{\rm Ia}) \ dt_{\rm Ia} } \ .
\label{eq:t_Ia}
\end{eqnarray}
Therefore, a more accurate estimate is obtained by the same
formulation but with a summation over galaxies satisfying $t_l \leq
\langle t_{\rm Ia} \rangle < t_u$. However, this integration diverges
when $t_{\rm Ia} \rightarrow 0$ if $\alpha \leq -1$.  The possibility
of a significant population of the prompt SNe Ia with $t_{\rm Ia}
\lesssim 0.1$ Gyr has been discussed in recent years (Scannapieco \&
Bildsten 2005; Mannucci et al. 2006; Sullivan et al. 2006; Aubourg et
al. 2007). Such a prompt population may also affect the estimate of
the delay time. Therefore, we separate the ``prompt'' population
(defined by $t_{\rm Ia} < t_p = 0.1$ Gyr) from the ``delayed''
(``tardy'') population ($t_{\rm Ia} \geq t_p$). Then, the DTD of the
delayed component can be estimated by a modified version of
eq. (\ref{eq:N_SN}),
\begin{equation}
N_{\rm Ia, exp} + N_{\rm pIa, exp} + N_{\rm CC, exp} = N_{\rm obs} \ , 
\end{equation}
where $N_{\rm Ia, exp}$ is now for the delayed population and the
integration of eq. (\ref{eq:t_Ia}) is from $t_{\rm Ia} = t_p$ to
$t_{\rm ga}$.  The expected number of the prompt Ia, $N_{\rm pIa,
  exp}$, can be calculated from SFR in a similar way to get $N_{\rm
  CC, exp}$, if the integrated DTD at $0 \leq t_{\rm Ia} \leq t_p$ is
given.

An obvious difficulty in this new formulation is that we need to know
the DTD itself beforehand to make a measurement of DTD, because
$\langle t_{\rm Ia} \rangle$ and $N_{\rm pIa, exp}$ depend on
$f_D(t_{\rm Ia})$.  In fact, this circularity problem can be solved
easily; first we estimate $f_D$ with an initial guess of $f_D(t_{\rm
  Ia})$, and then we can iterationally repeat this process with the
new estimate of $f_D (t_{\rm Ia})$.  The first DTD estimate with a
constant $f_D$ prior (open squares in Fig. \ref{fig:dtd_obs}) is well
described by a power-law ($\propto {t_{\rm Ia}}^\alpha$) at $t_{\rm
  Ia} \ge t_p$, and hence we assume this form of $f_D$ for the delayed
population at the later iterations. For the prompt population, we
assume that $f_D$ is constant at $t_{\rm Ia} \leq t_p$ and it is
continuously connected to the delayed component at $t_{\rm Ia} = t_p$,
for every iteration. This is reasonable from the DTD predictions by
the stellar evolution theory (section \ref{section:models}), and the
effect of this assumption on the DTD measurement will be examined in
section \ref{section:systematics}.

We get sufficiently convergent results by just a few iterations of
this procedure, and they are shown by filled squares in
Fig. \ref{fig:dtd_obs} as our final results.  Since we selected old
galaxies, $\langle t_* \rangle$ is already a good estimate of $\langle
t_{\rm Ia} \rangle$, and this is why the correction is not large. The
expected number of the prompt population, $N_{\rm pIa, exp}$, is 3.6
for the entire range of $t_{\rm Ia} = $ 0.1--8 Gyr, which is just
5.5\% of the 65 SN candidates.  The derived DTD, as well as important
quantities such as $N_{\rm pIa, exp}$, $N_{\rm CC, exp}$, and $N_{\rm
  obs}$, are summarized in Table \ref{table:baseline} for the six time
bins in $t_{\rm Ia}$ = 0.1--8 Gyr.  The best fit power-law to the
measured DTD is $f_D({\rm 1 \ Gyr}) = 0.55^{+0.12}_{-0.11} \ {\rm
  century^{-1}} (10^{10} L_{K,0,\odot})^{-1}$ and $\alpha =
-1.08^{+0.15}_{-0.15}$.  Here, we included the data point at $t_{\rm
  Ia} = 11$ Gyr in the fit.

\section{Examination of Systematic Uncertainties}
\label{section:systematics}

\subsection{Are the SN candidates really SNe Ia?}

In our statistical estimation of DTD, it is not necessary to prove that
all of the SN candidates are SNe Ia.  Since we have already shown that
the selected variable objects are tracing the light profile of host
galaxies, the majority of them must be SNe Ia or CC SNe. What we need to
demonstrate is, then, that about 82\% of the 65 candidates are actually
SNe Ia, as inferred from the estimate of $N_{\rm CC, exp} = 11.4$.

We first examine the $i'$ band variability magnitude $m_{\rm var, \it
  i'}$ of the 65 SN candidates, and compare with the brightest
magnitude $m_{\rm min, \it i'}$ that is possible for SNe Ia
corresponding to the peak luminosity.  We calculated this magnitude
for each SN candidate assuming the standard type Ia light curve with
the mean peak $B$ magnitude used in Oda \& Totani (2005). We find that
most (61/65) of the SN candidates have $m_{\rm var, \it i'} > m_{\min,
  \it i'}$ as expected, and all objects satisfy $m_{\rm var, \it i'}
\geq m_{\min, i'} - 0.53$.  The four objects with $m_{\rm var, \it i'}
< m_{\min, i'}$ are reasonable considering the dispersion of the peak
$B$ magnitude of SNe Ia ($\sim$0.4 mag). On the other hand, about half
of the SN candidates have the variability magnitudes brighter than
$m_{\rm min, \it i'} + 1$, and such variability cannot be explained by
CC SNe, because most of them are fainter than SNe Ia by 1--2
magnitudes (see e.g., Oda \& Totani 2005).

\begin{figure*}
\begin{center}
\includegraphics[width=70mm,angle=-90]{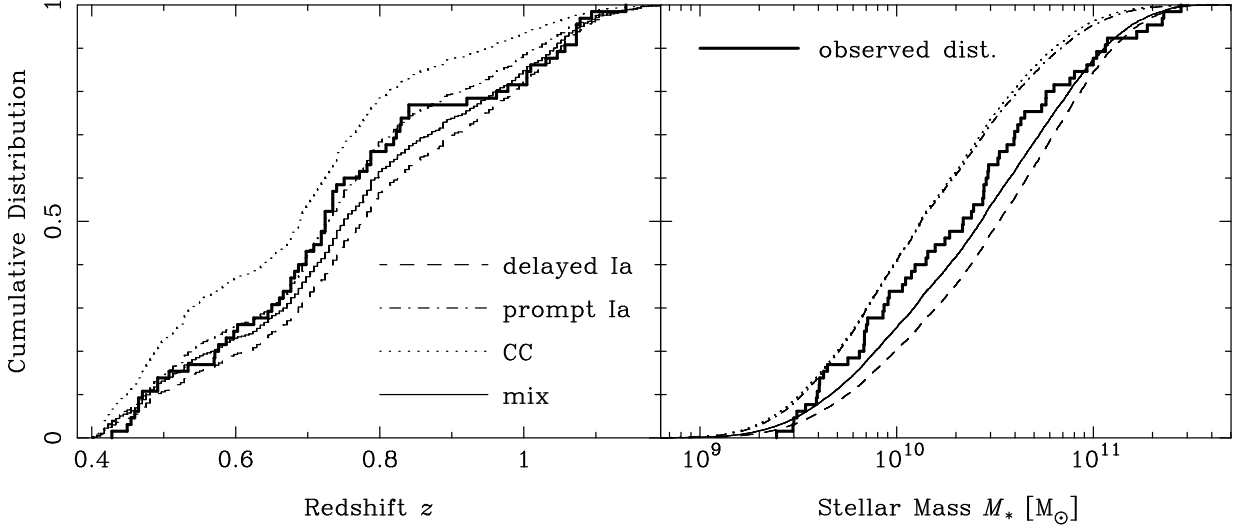} 
\end{center}
\caption{The cumulative distributions of redshift and stellar mass of
host galaxies for the SN candidates.  The thick solid lines are the
observed distributions of the 65 SN candidates. The dashed and
dot-dashed curves are the expected distributions for the delayed ($t_{\rm Ia}
\ge$ 0.1 Gyr) and prompt ($t_{\rm Ia} <$ 0.1 Gyr) components of SNe Ia,
respectively, and the dotted curve is that for CC SNe. The thin solid
curve is our best guess, which is the sum of the three components with
the expected numbers in the baseline analysis: $N_{\rm exp} = 45.2$
(delayed Ia), 3.6 (prompt Ia), and 11.4 (CC).  The results of the
Kolmogorov-Smirnov test for these distributions are given in Table
\ref{table:KS}.}  \label{fig:zM_dist}.
\end{figure*}

We already made the DTD estimates, and we can predict the expected
distribution of redshifts and stellar masses of host galaxies for the SN
candidates, because we can calculate the expected number of SNe Ia in
each galaxy. Comparison of these expected distributions with those
observed provides us with an important consistency check of our DTD
estimates.  Figure \ref{fig:zM_dist} shows such comparisons, and the
predictions based on our final DTD estimates are in nice agreement with
the data.  In the calculation of the expected redshift and stellar mass
distributions, the light curve information of SNe is included through
the visibility time. Therefore, we do not expect such an agreement, if
our estimate of Ia/CC ratio is wrong or there is a significant
contamination from any non-SN objects. In fact, the predicted
distributions are in serious contradiction with the data if we assume
that all the SN candidates are CC SNe. The agreement between the
expected and observed distributions are quantitatively tested by the
Kolmogorov-Smirnov test, and the results are given in Table
\ref{table:KS}. Acceptable fits are obtained both for the redshift and
stellar mass distributions only when we make an appropriate mix of the
delayed/prompt Ia and CC SNe with the relative proportions predicted by
our DTD estimates: $N_{\rm exp}$ = 45.2 (delayed Ia), 3.6 (prompt Ia), and
11.4 (CC).  These results then give a strong support to the reliability
of our DTD estimates.

\begin{table}
\caption{ The results of the Kolmogorov-Smirnov tests for the redshift
and host galaxy stellar mass distributions of the SN candidates.  The
chance probabilities of getting the observed distributions are
shown. (See Fig. \ref{fig:zM_dist} for the graphical presentations of
these distributions.)  The columns 2--4 show the results when all SNe
are assumed to be delayed SNe Ia, prompt SNe Ia, and CC SNe,
respectively. The last column shows the results when the three
populations are mixed with the expected numbers (45.2, 3.6, and 11.4) in
our baseline analysis.}
\begin{center}
\begin{tabular}{ccccc}
\hline
\hline
   & delayed Ia & prompt Ia & \ CC \ & \ mix \ \\
\hline
 redshift              & 0.07 & 0.81 & 0.05 & 0.40 \\
 stellar mass          & 0.05 & 0.07 & 0.04 & 0.48 \\  
\hline
\hline
\end{tabular}
\end{center}
\label{table:KS}
\end{table}

It should also be noted that we selected all variable objects
associated with the old galaxies with a significant offset from the
nuclei.  Only 2 objects were removed by AGN-like variability, which is
a negligible number compared with the final 65 candidates. Therefore,
our DTD measurement gives a conservative upper limits to the true DTD,
and the derived DTD would always be overestimates if there was
any unknown contamination from non-SN objects.

\begin{figure*}
\begin{center}
\includegraphics[width=100mm,angle=-90]{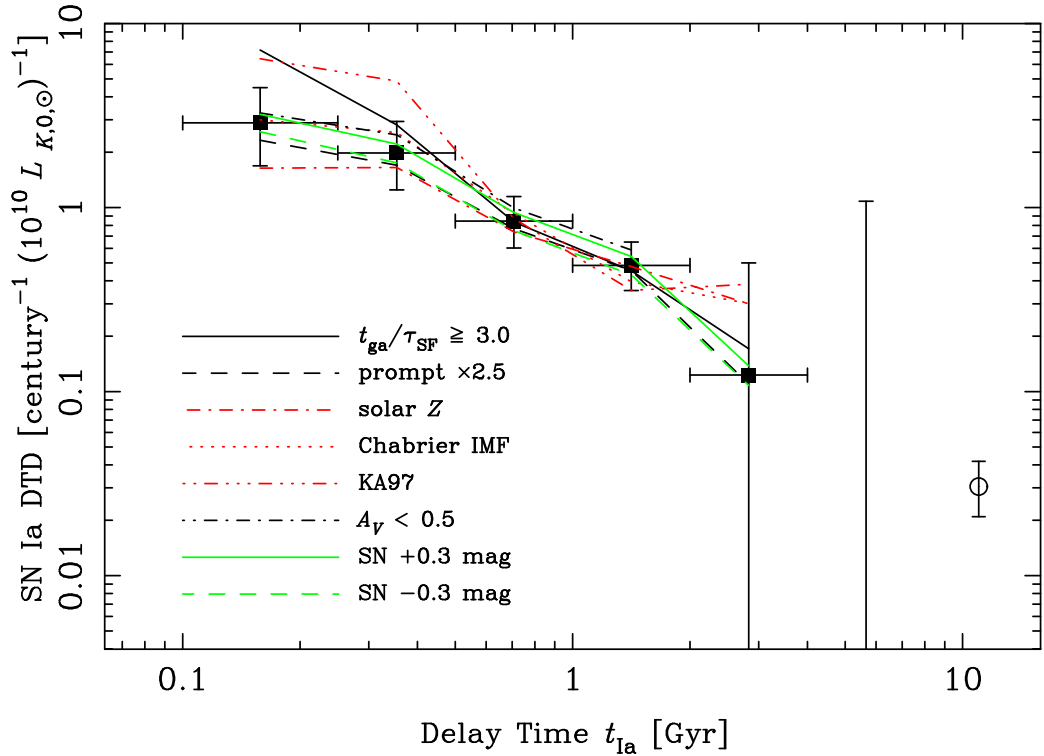}
\end{center}
\caption{ The SN Ia DTD $f_D(t_{\rm Ia})$ estimated by different
  prescriptions from the baseline analysis are shown by lines. The
  data points are for the baseline analysis, which are the same as
  those in Fig. \ref{fig:dtd_obs}.  The labels for the lines are the
  same as in Table \ref{table:systematics}, and see this table and
  main text for explanations.  }
\label{fig:dtd_syst}
\end{figure*}

Finally, we repeated our DTD estimates with a more stringent criteria
for the old galaxies, $t_{\rm ga} / \tau_{\rm SF} \geq 3.0$, to select
more efficiently SNe Ia rather than CC SNe.  In this case, the number
of the SN candidates is reduced to 44, but now $N_{\rm CC, exp} = 5.4$
and $N_{\rm pIa, exp} = 1.6$, and hence the expected Ia
(delayed+prompt) fraction is increased to 88\% from 82\% in the
baseline analysis.  The DTD estimates and the power-law fit in this
case are shown in Table \ref{table:systematics} and
Fig. \ref{fig:dtd_syst}, and the difference from the baseline results
is within the statistical uncertainties.  It should be noted that the
number of the SN candidates in the shortest bin (0.1--0.25 Gyr) is now
only 5, because the possible minimum of mean stellar age is increased to
0.22 Gyr from 0.16 Gyr in the baseline analysis.  Therefore, although
the DTD value in the shortest time bin shows a larger deviation from
the baseline analysis than in other time bins, the statistical
uncertainty is quite large.

\begin{table*}
  \caption{ Examination of systematic errors in the DTD measurements.
    Various DTD results are shown when the analysis method is changed from
    the baseline analysis.  The last two columns show the best-fit power-law
    DTD, $f_D(t_{\rm Ia}) = f_{D, \rm  1Gyr} (t_{\rm Ia}/{\rm 1 \
      Gyr})^{\alpha}$.  The ``$t_{\rm Ia} = \langle t_* \rangle$'' results are
    obtained by estimating $t_{\rm Ia}$ simply by mean stellar age $\langle
    t_* \rangle$ of the host galaxies (open squares in Fig. 
    \ref{fig:dtd_obs}). The ``$t_{\rm ga}/\tau_{\rm SF} \geq 3.0$''
    results are obtained when a more strict criterion of the old galaxies is
    applied than the baseline analysis.  The ``prompt $\times$2.5'' results
    are obtained when the fraction of the prompt Ia population is increased
    by a factor of 2.5 from the baseline analysis.  The ``solar $Z$'',
    ``Chabrier IMF'' and ``KA97'' results are obtained using stellar 
    age and mass 
    estimates with a different metallicity, a different IMF, and a different
    stellar population synthesis model from the baseline analysis,
    respectively. The ``$A_V < 0.5$'' results are obtained with a
    more strict cut about the dust extinction of host galaxies. 
    The ``SN $\pm$0.3 mag'' results are obtained with the SN
    Ia light curve $\pm 0.3$ mag fainter/brighter than in the baseline
    analysis. All errors are in statistical 1$\sigma$.}  \footnotesize
\begin{center} 
\begin{tabular}{cccccccccc}
\hline
\hline
& \multicolumn{6}{c}{$f_D(t_{\rm Ia})$
[century$^{-1}(10^{10} L_{K, 0, \odot})^{-1}$]
in $t_{\rm Ia}$ bins [Gyr]} & &  \\ 
\cline{2-7} 
Analysis & 0.1--0.25 & 0.25--0.5 & 0.5--1.0 & 1.0--2.0 & 2.0--4.0
& 4.0--8.0 & \ \ \ \ $f_{D, \rm 1Gyr}$ \ \ \ \ & $\alpha$ \\
\hline
baseline & $2.89^{+1.60}_{-1.20}$ & $1.98^{+0.96}_{-0.73}$ &  
$0.84^{+0.30}_{-0.24}$ & $0.49^{+0.17}_{-0.13}$ & 
$0.12^{+0.38}_{-0.12}$ & $0.00^{+1.08}_{-0.00}$ & 
$0.55^{+0.12}_{-0.11}$ & $-1.08^{+0.15}_{-0.15}$ \\
$t_{\rm Ia} = \langle t_* \rangle$ & 
$3.12^{+1.64}_{-1.21}$ & $2.87^{+1.26}_{-0.96}$ &  
$0.89^{+0.36}_{-0.28}$ & $0.52^{+0.16}_{-0.13}$ & 
$0.56^{+0.42}_{-0.27}$ & $0.00^{+0.47}_{-0.00}$ & 
$0.63^{+0.14}_{-0.12}$ & $-1.09^{+0.15}_{-0.15}$ \\
$t_{\rm ga}/\tau_{\rm SF} \geq 3.0$ & 
$7.18^{+5.65}_{-3.61}$ & $2.81^{+1.76}_{-1.23}$ &  
$0.84^{+0.33}_{-0.26}$ & $0.45^{+0.17}_{-0.13}$ & 
$0.17^{+0.47}_{-0.17}$ & $0.00^{+1.19}_{-0.00}$ & 
$0.65^{+0.17}_{-0.14}$ & $-1.23^{+0.18}_{-0.17}$ \\
prompt $\times$2.5 & $2.32^{+1.60}_{-1.20}$ & $1.70^{+0.96}_{-0.73}$ &  
$0.77^{+0.30}_{-0.24}$ & $0.46^{+0.17}_{-0.13}$ & 
$0.11^{+0.38}_{-0.11}$ & $0.00^{+1.07}_{-0.00}$ & 
$0.50^{+0.12}_{-0.12}$ & $-1.05^{+0.16}_{-0.15}$ \\
solar $Z$ & $1.64^{+0.81}_{-0.62}$ & $1.65^{+0.84}_{-0.63}$ &  
$0.74^{+0.40}_{-0.30}$ & $0.48^{+0.16}_{-0.13}$ & 
$0.30^{+0.19}_{-0.13}$ & $0.14^{+0.38}_{-0.14}$ & 
$0.47^{+0.11}_{-0.10}$ & $-0.92^{+0.15}_{-0.13}$ \\
Chabrier IMF & $3.00^{+1.65}_{-1.23}$ & $2.56^{+1.15}_{-0.89}$ &  
$0.93^{+0.34}_{-0.27}$ & $0.40^{+0.17}_{-0.13}$ & 
$0.30^{+0.48}_{-0.23}$ & $0.00^{+0.84}_{-0.00}$ & 
$0.57^{+0.13}_{-0.12}$ & $-1.11^{+0.15}_{-0.15}$ \\
KA97 & $6.43^{+2.65}_{-2.08}$ & $4.87^{+1.43}_{-1.18}$ &  
$0.87^{+0.55}_{-0.39}$ & $0.36^{+0.25}_{-0.17}$ & 
$0.38^{+0.33}_{-0.20}$ & $0.43^{+1.02}_{-0.37}$ & 
$0.79^{+0.19}_{-0.16}$ & $-1.27^{+0.14}_{-0.14}$ \\
$A_V < 0.5$ & $3.27^{+2.88}_{-1.97}$ & $2.48^{+1.57}_{-1.15}$ &  
$1.00^{+0.47}_{-0.36}$ & $0.59^{+0.20}_{-0.16}$ & 
$0.00^{+0.35}_{-0.00}$ & $0.00^{+1.35}_{-0.00}$ & 
$0.64^{+0.17}_{-0.16}$ & $-1.16^{+0.17}_{-0.16}$ \\
SN $+0.3$ mag & $3.21^{+1.78}_{-1.34}$ & $2.21^{+1.07}_{-0.82}$ &  
$0.94^{+0.34}_{-0.27}$ & $0.54^{+0.18}_{-0.15}$ & 
$0.14^{+0.42}_{-0.14}$ & $0.00^{+1.22}_{-0.00}$ & 
$0.60^{+0.13}_{-0.12}$ & $-1.11^{+0.15}_{-0.15}$ \\
SN $-0.3$ mag & $2.58^{+1.44}_{-1.07}$ & $1.75^{+0.85}_{-0.65}$ &  
$0.75^{+0.27}_{-0.21}$ & $0.43^{+0.15}_{-0.12}$ & 
$0.11^{+0.33}_{-0.11}$ & $0.00^{+0.96}_{-0.00}$ & 
$0.50^{+0.11}_{-0.10}$ & $-1.05^{+0.15}_{-0.15}$ \\
\hline
\hline
\end{tabular}
\end{center}
\label{table:systematics}
\end{table*}

\subsection{Estimates of $\langle t_{\rm Ia} \rangle$ and contribution
  from the prompt Ia population}
\label{section:syst_tIa}

In our analysis, the delay time $t_{\rm Ia}$ of each SN candidate has
been estimated simply by the expectation value of $\langle t_{\rm Ia}
\rangle$ from the star formation history of the host galaxy. We have
already shown that the difference between $\langle t_{\rm Ia} \rangle$
and the mean stellar age $\langle t_* \rangle$ is small and does not
seriously affect the final results.  However, the probability
distribution of $t_{\rm Ia}$ should have some dispersion around $\langle
t_{\rm Ia} \rangle$, and if this dispersion is larger than the bin width
of the DTD estimate, it could affect the results. The dispersion around
$\langle t_{\rm Ia} \rangle$, $\sigma_{t_{\rm Ia}} = (\langle t_{\rm
Ia}^2 \rangle - \langle t_{\rm Ia} \rangle^2)^{1/2}$ is therefore
calculated for each host galaxy in a similar way to eq. (\ref{eq:t_Ia}),
and their average in a $t_{\rm Ia}$ bin is shown in Table
\ref{table:baseline}. The dispersions are typically about half of the
bin widths, and hence the binning size of our DTD estimates is
appropriate.

In our baseline analysis, the amount of prompt SNe Ia is calculated by
assuming a constant $f_D(t_{\rm Ia})$ at $t_{\rm Ia} \leq t_p$, and in
this case the fraction of the prompt Ia is about 20\% of all SNe Ia
when the DTD is integrated over a range of $t_{\rm Ia} = 0$--11
Gyr. However, since we selected old galaxies, $N_{\rm pIa, exp} = 3.6$
is just 5.5\% of our 65 SN candidates. The expected fraction of prompt
SNe Ia is the largest in the shortest delay time bin (0.1--0.25 Gyr),
but it is still less than 10\% (1.0/12, see Table
\ref{table:baseline}). Therefore, our result will not be seriously
affected by changing the prompt fraction, unless we assume an
extremely higher prompt Ia rate (e.g., by a factor of about 10) than
that assumed in the baseline analysis. We show the results of the DTD
estimate when the prompt Ia population is enhanced by a factor of 2.5
from our baseline analysis in Table \ref{table:systematics} and
Fig. \ref{fig:dtd_syst}. In this case, the prompt fraction in all SNe
Ia integrated over $t_{\rm Ia} = $0--11 Gyr becomes $\sim 50$\%, which
is a typical value discussed in recent papers about prompt SNe Ia
(e.g., Mannucci et al. 2006; Sullivan et al. 2006). Our DTD estimate
for the delayed SNe Ia is not seriously affected.

\subsection{Systematic Uncertainties in Stellar Age and Mass}
\label{section:syst_age_mass}

The estimates of stellar mass and age of the host galaxies are crucial
in our DTD estimates.  Since we selected old or passively evolving
galaxies in a relatively bright magnitude range of $m_{i'} \leq 24$,
the estimates are expected to be easier than general studies of
high-$z$ galaxies.  However, these estimates may sensitively depend on
e.g. the SF history, IMF, and metallicity assumed in the photometric
redshift calculation. Therefore the systematic uncertainties must be
carefully examined.

We test several age and mass estimations with different prescriptions
of the photometric redshift calculation as follows.  We first try the
two cases of changing metallicity into the solar abundance ($Z=0.02$)
and changing IMF into the Chabrier (2003) IMF, from the baseline
analysis. The choice of the supersolar metallicity ($Z = 0.05$) in the
baseline analysis is motivated by the metallicity estimates of local
early-type galaxies (Barber et al. 2007; Jimenez et al. 2007), as
mentioned in \S \ref{section:photo-z}. Recent observations of the
mass-metallicity relation at various redshifts give a further support
to this choice. According to the mass-metallicity relation of Savaglio
et al. (2005) measured for galaxies at a similar redshift range ($0.4
\leq z \leq 1.0$) to our sample, metallicity of the old galaxies
selected in this work is expected to be larger than the solar value,
even when our definition of the stellar mass (integration of SFR) is
taken into account. Therefore our examination in the metallicity range
of $Z$ = (1--2.5) $Z_\odot$ is reasonable to check the systematic
uncertainties in DTD.

Furthermore, we used a completely different population synthesis model
by Kodama \& Arimoto (1997, hereafter KA97), based on an independent
stellar spectrum library. We used the $M_V= -21.98$ model in the
metallicity-sequence KA97 models as a template. In this model, a gas
infall and galactic wind are taken into account rather than a simple
exponential SFR evolution, and the chemical evolution is solved in a
self consistent manner. This is a model for a local elliptical galaxy,
and the star formation time scale is about $\tau_{\rm SF} \sim$ 0.1
Gyr.  We have only one template for the star formation history, and
hence the age estimate is likely to be less accurate than the baseline
analysis, but we can check the dependence on different stellar
population synthesis models as well as on the number of used SF
history templates.

\begin{figure*}
\begin{center}
\includegraphics[width=60mm,angle=-90]{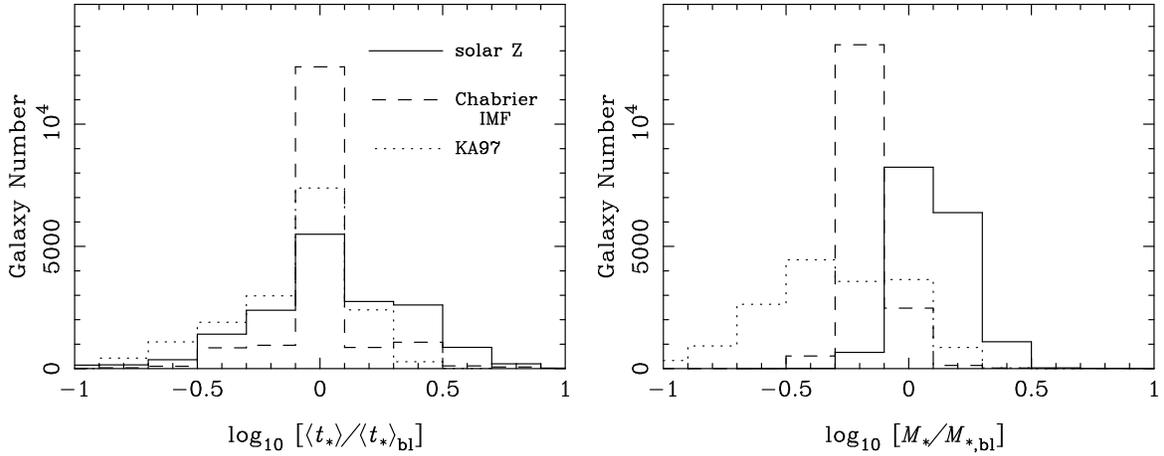} 
\end{center}
\caption{The histograms of the mean stellar ages $\langle t_* \rangle$
(left) and stellar mass $M_*$ (right)
of the old galaxies
estimated by alternative photometric redshift calculations to the
baseline analysis. They are shown in the ratios to those in the
baseline analysis. The solid and dashed lines are for the cases of
changing to the solar metallicity ($Z = 0.02$) and to the Chabrier IMF,
respectively. The dotted line is for the estimates based on a
completely different model (KA97) of stellar population synthesis.}
\label{fig:age_hist}
\end{figure*}

In Fig. \ref{fig:age_hist}, we show the distribution of $\langle t_*
\rangle / \langle t_* \rangle_{\rm bl}$ and $M_* / M_{*, \rm bl}$ of
the old galaxies, i.e., the ratios of $\langle t_* \rangle$ and $M_*$
estimated by different prescriptions to those in our baseline
analysis.  The logarithmic means and standard deviations of these
quantities are tabulated in Table \ref{table:mean_sig_var}.  It can be
seen that the means of $\langle t_* \rangle$ are not significantly
changed, and the typical dispersion of $\langle t_* \rangle / \langle
t_* \rangle_{\rm bl}$ is about a factor of two. This means that the
bin width of our DTD measurements is not unreasonably small compared
with the systematic uncertainties.  As for stellar masses, systematic
offsets of mean values of $M_* / M_{*, \rm bl}$ from unity can be
seen.  However, for the solar $Z$ and the Chabrier IMF cases, these
offsets are considerably reduced when we consider the $K$-band
luminosity at the age of 11 Gyr ($L_{K,0}$) as the mass indicator. The
mass-to-light ratios are $M_*/L_{K,0} = 2.0$ and 1.4
$[M_\odot/L_{K,\odot}]$ for the solar $Z$ and the Chabrier IMF cases,
respectively (1.8 for the baseline analysis), 
and hence the effective offsets in $\log_{10} L_{K,0}$
now become 0.056 and $-$0.043, respectively, i.e., less than $\sim
10$\% difference in DTD estimates. This demonstrates the merit of
using $L_{K,0}$ as the mass indicator to reduce the uncertainties
about stellar mass estimates. On the other hand, $M_*/L_{K,0} = 2.1 \
[M_\odot/L_{K,\odot}]$ for the KA97 model and it cannot compensate the
large offset of $M_*/M_{*, \rm bl}$.  This is most likely a result of
the limited SF history template in the estimates using the KA97 model
(only 1 template corresponding to $\tau_{\rm SF} \sim 0.1$ Gyr). It is
expected that the age estimates with a single small value of
$\tau_{\rm SF}$ tend to be underestimate of the true age, and this
trend is in fact seen in the left panel of Fig.
\ref{fig:age_hist}. The underestimate in stellar age would result in
an underestimate in the stellar mass.

To examine the sensitivity of our DTD estimates to these uncertainties
of stellar ages and masses, we repeated the DTD estimates but using
$M_*$, $(M_*/L_{K,0})$, and $\langle t_{\rm Ia} \rangle$
calculated with the different prescriptions, and the results are shown
in Table \ref{table:systematics} and Fig. \ref{fig:dtd_syst}.  The
change in the DTD is within the statistical 1$\sigma$ errors of the
baseline results for most of the data points. The KA97 model case shows
a large deviation from the baseline analysis at $t_{\rm Ia} \lesssim
0.5$ Gyr, compared with the other cases.  This is probably due to the
age underestimates by a single SF template as discussed above, and we
consider that the baseline results are more reliable than those using
the KA97 model. The data points at $t_{\rm Ia}$ = 0.5--2 Gyr are quite
robust against these tests.

\begin{figure*}
\begin{center}
\includegraphics[width=60mm,angle=-90]{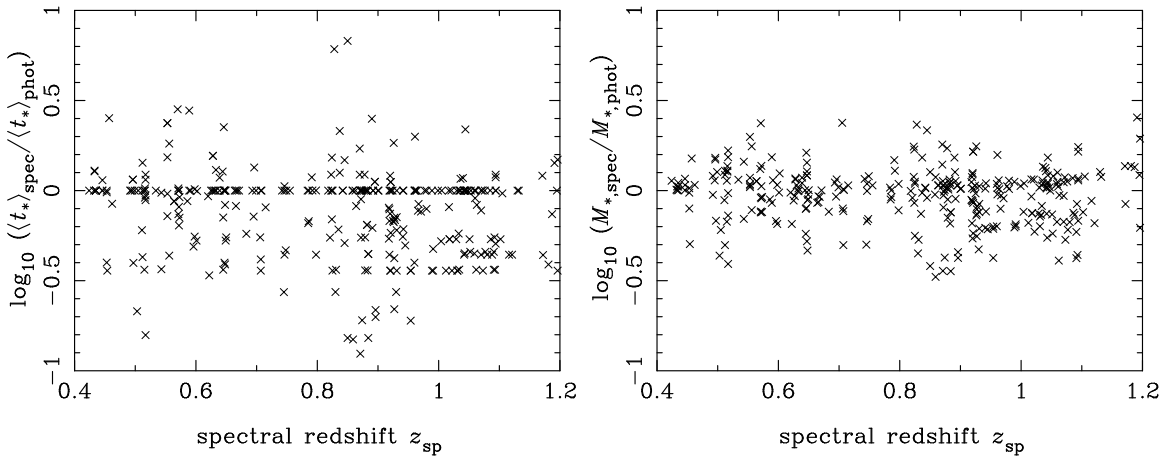} 
\end{center}
\caption{The ``spectroscopic'' mean stellar ages ($\langle t_*
  \rangle$) and stellar masses ($M_*$) versus spectroscopic redshifts,
  for the old galaxies with available spectroscopic data.  See the
  main text for the definition of these quantities. They are given as
  the ratios to those by the normal photometric redshift calculations.
}
\label{fig:age_mass_spec}
\end{figure*}

Furthermore, we calculate $\langle t_* \rangle_{\rm spec}$ and $M_{*,
  \rm spec}$ for the old galaxies with spectroscopic redshifts, to
examine the systematic uncertainties in stellar age and mass coming
from the uncertainty about redshift. These are calculated from the
same photometric redshift calculations, but fixing the redshift at the
value known by spectroscopic observations. The results are shown in
Fig. \ref{fig:age_mass_spec}, as the ratios to $\langle t_*
\rangle_{\rm phot}$ and $M_{*, \rm phot}$ obtained by the normal
photometric redshift calculations (i.e., redshift treated as a free
parameter). The logarithmic mean and the standard deviation of these
ratios are shown in Table \ref{table:mean_sig_var}. Many galaxies have
exactly the same value of $\langle t_* \rangle_{\rm spec} = \langle
t_* \rangle_{\rm phot}$ on the age grids of the {\it hyperz} code, and
73\% of the old galaxies have $\langle t_* \rangle_{\rm spec}$ within
a factor of two from $\langle t_* \rangle_{\rm phot}$.  It seems that
these uncertainties are not greater than the others about the
photometric age/mass estimates discussed above.

\begin{table*}
\caption{The logarithmic mean and standard deviation of the ratios of
mean stellar age and stellar mass to those in the baseline analysis,
for alternative photometric redshift calculations using
the solar metallicity, the Chabrier IMF, the KA97 galaxy evolution model,
and fixing redshifts at the spectroscopic values.}
\begin{center}
\begin{tabular}{cccccc}
\hline  \hline
 & & solar $Z$ & Chabrier IMF & KA97 & spec \\
\hline
$\log_{10} [\langle t_*\rangle / \langle t_{*} \rangle_{\rm bl}]^*$ & mean
& 0.04 & 0.004 & $-$0.09 & $-$0.11 \\
& sigma & 0.35 & 0.23 & 0.27 & 0.25 \\
$\log_{10} [M_* / M_{*, \rm bl}]^*$ & mean
& 0.10 & $-$0.16 & $-$0.30 &  $-$0.02 \\
& sigma & 0.15 & 0.17 & 0.28 & 0.15  \\
\hline \hline
\end{tabular}
\end{center}
{ \footnotesize
$^*$These quantities should be replaced by 
$\langle t_{*} \rangle_{\rm spec}/\langle t_* \rangle_{\rm phot}$ and
$M_{*, \rm spec} / M_{*, \rm phot}$ for the ``spec'' column. 
}
\label{table:mean_sig_var}
\end{table*}

A caveat in our analysis is that the variety of SF history used in
photo-$z$ calculation is limited. It is computationally impractical to
increase the number of SF history templates by a large factor from
that used in this work (eight). However, the dependence on different
SF histories has partially been tested by the KA97 model case as
described above. We also note that, since we selected old galaxies,
the estimates of mean stellar age are expected to be rather
insensitive to assumed SF histories.  What is important for DTD
estimates is the mean stellar ages rather than the details of SF
history, provided that the dispersion of stellar age around the mean
is not greater than the bin width of DTD measurements. This has
already been checked by the calculations of $\langle \sigma_{t_{\rm
    Ia}} \rangle$ in section \ref{section:syst_tIa}.

A considerable fraction of the old galaxies have $A_V > 0.5$ in spite
of low star formation activity, as shown in Fig. \ref{fig:i_hist}. It
might be an artifact due to the uncertainties in $A_V$ estimates, but
in that case it would affect the age estimates. Therefore we removed
the old galaxies with $A_V > 0.5$ and repeated the DTD calculations,
which are shown in Table \ref{table:systematics} and
Fig. \ref{fig:dtd_syst}. The number of the SN candidates is reduced to
48, but the change from the baseline analysis is not significant. The
effect of dust extinction on supernova luminosities will be discussed
in the next subsection.

We have removed the nuclear region of host galaxies within 1.9 pix
radius in our SN candidate search, and it might induce some biases in
the age estimates if the removed nuclear regions have considerably
different stellar age from the outer regions. The mean fraction of
$i'$ flux within the nuclear region of the old galaxies is 35\% of the
total isophotal flux, and hence the outer regions are more dominant
than the nuclear regions in the photometric redshift calculation.
Figure \ref{fig:images} indicates that most of the old galaxies have
elliptical morphologies, and they are likely to be the ancestors of
the present-day elliptical galaxies (Yamada et al. 2005). It is known
that the radial color gradient of local elliptical galaxies is
explained by metallicity gradient rather than age gradient (Tamura et
al. 2000).  The typical metallicity gradient is $d\log Z / d\log r
\sim 0.3$ (Kobayashi \& Arimoto 1999), and the effect of metallicity
gradient has already been tested, at least partially, by the above
examination of the age dependence on metallicity.

\subsection{Systematic Uncertainties in Supernova Luminosity}

The light curves of SNe are essential for calculations of the
visibility time and hence for a rate study. We used the standard light
curves of various supernova types as in Oda \& Totani (2005). However, we
selected SNe Ia in galaxies with old stellar population, and a
systematic trend that SNe Ia in old stellar population are fainter
than those in star forming galaxies by a peak magnitude difference of
$\Delta M_B \sim 0.3$ has been known (Gallagher et al. 2005;
Sullivan et al. 2006). 

Extinction in host galaxies may also change the apparent brightness of
SNe, and hence the effective luminosity.  We assumed a low extinction
of $E(B-V) = 0.05$ (or $A_V = 0.155$ for the standard Milky Way
extinction curve) for SNe Ia, which is reasonable for old
galaxies. However, this value is smaller than $\langle A_V \rangle =
0.31$ of the old galaxies estimated by the photo-$z$ code, though this
value should not naively be taken as the extinction of SN flux, as
mentioned in section \ref{section:formulations}. It should also be
noted that the distribution of old galaxy $A_V$ is strongly peaked at
$A_V = 0$ (Fig. \ref{fig:i_hist}), and the uncertainty in $A_V$
estimate may have resulted in a larger $\langle A_V \rangle$ than the
real value.

To examine the sensitivity of the DTD estimates to the effective
changes of SN luminosity by these effects, we repeated the DTD
estimates with the SN Ia light curve shifted by $\pm 0.3$ mag, and the
results are shown in Table \ref{table:systematics} and
Fig. \ref{fig:dtd_syst}. The difference from the baseline analysis is
small, and the uncertainty about the SN luminosity or host galaxy
extinction is unlikely to change our main conclusions.

\subsection{Uncertainties in the Comparison with the SN Ia Rate of
Local Elliptical Galaxies}
\label{section:local_data}

We assumed 11 Gyr for the mean age of the local elliptical galaxies
and hence for the data of Mannucci et al. (2005).  We have chosen this
value as a mean value of the luminosity-weighted stellar ages of the
early type galaxies at $0.05 \leq z \leq 0.2$ estimated by Jimenez et
al. (2007), based on the MOPED fits to the SDSS spectra.  A slightly
smaller but similar age ($\sim$ 9 Gyr) is obtained for the SDSS
luminous red galaxies at $0.15 \leq z \leq 0.4$ (Barber et al. 2007),
when the look back time to the redshift of these galaxies is
corrected.  The age dispersion around the mean for these early-type
galaxies is about 2 Gyr, both for the samples of Jimenez et al. and
Barber et al.  A systematic uncertainty of $\sim$ 2 Gyr in the 11 Gyr
data point does not significantly affect the conclusion that the DTD
is well described by a power-law.  \footnote{Very recently, Gallagher
  et al. (2008) reported the age and metallicity estimates for the
  elliptical host galaxies of the local SNe Ia, and the ages estimated
  by them have a considerably wider distribution than those of Jimenez
  et al. (2007) and Barber et al (2007), including a significant
  fraction of young galaxies with ages less than 5 Gyr. There is no
  difference in the properties of the SN Ia host galaxies and field
  elliptical galaxies (Gallagher et al. 2008), indicating that the
  age difference comes from some systematic uncertainties in the age
  estimates.  The origin of the difference is not clear to us, but it
  should be noted that the age estimate by Gallagher et al. (2008) is
  based on H$\beta$, Fe$\lambda5270$ and Fe$\lambda$4383 indices with
  the spectral library including only single-burst stellar population
  (SSP).  In such analyses, the age estimates are easily affected by a
  small amount of recent star formation, and they stated in their
  paper that their age estimates should be regarded as lower limits to
  the true ages.  Their figure 7 shows a strong correlation between
  the ages and metallicities especially for galaxies younger than 5
  Gyr, indicating an effect of the age-metallicity degeneracy in the
  fittings.  The estimates by Jimenez et al. (2007) are based on full
  spectral information including continuum as well as absorption line
  indices, allowing any SF history in 11 time bins.}

In the SN Ia rate estimate for the local elliptical galaxies (Mannucci
et al. 2005), very faint SNe Ia with the stretch factor $s < 0.8$ are
included, reflecting the observed trend that SNe Ia found in old
galaxies are subluminous on average (Gallagher et al. 2005; Sullivan
et al. 2006).  On the other hand, such subluminous SNe Ia are hardly
detected in flux-limited high-$z$ supernova searches [see, e.g.,
Fig. 11 of Sullivan et al. (2006)]. If the fraction of subluminous SNe
Ia does not evolve with the delay time, we should correct our DTD data
points upward to make an unbiased comparison with the data of Mannucci
et al. (2005). However, the fraction of $s < 0.8$ SNe Ia in local
elliptical galaxies is $\sim$ 30\% [see again Fig. 11 of Sullivan et
al. (2006)], and it does not significantly affect the main conclusions
of this work. Furthermore, since our SN candidate sample is at high
redshift, it mainly probes shorter delay times of $t_{\rm Ia}
\lesssim$ 2 Gyr, and there may not be subluminous SNe Ia with such
short delay times. In this case, the correction discussed here 
is not necessary.

\section{Implications for the SN Ia Progenitor}
\label{section:models}

\subsection{the Double-Degenerate Scenario}

\begin{figure*}
\begin{center}
\includegraphics[width=100mm,angle=-90]{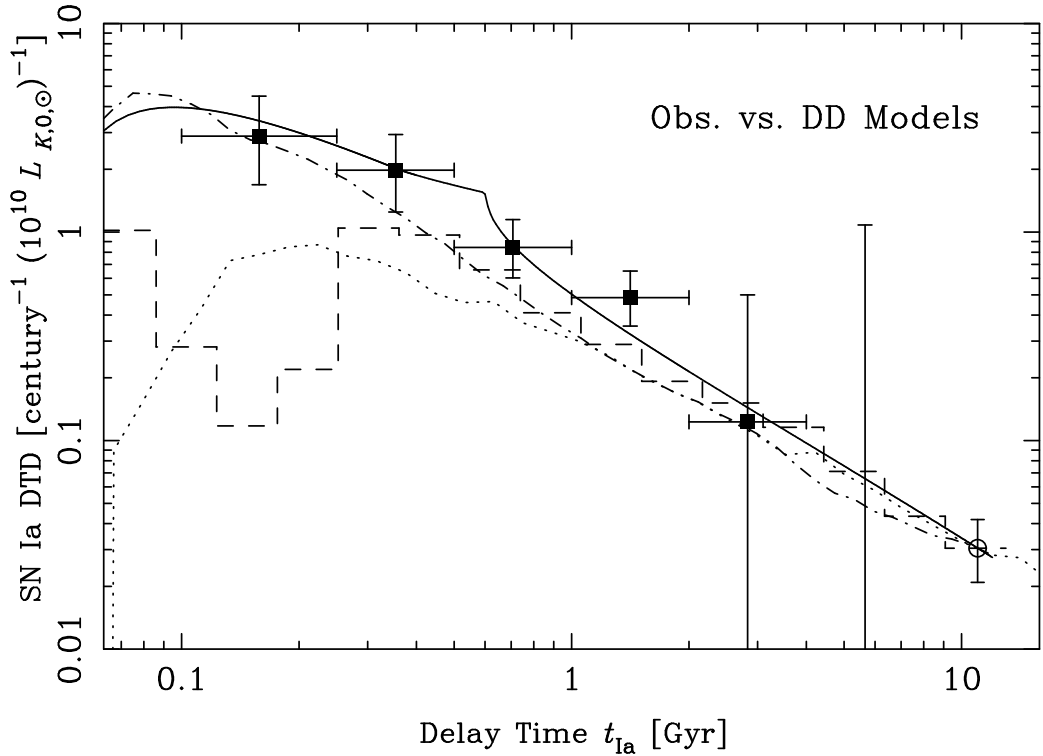}
\end{center}
\caption{The observed SN Ia DTD $f_D(t_{\rm Ia})$ compared with the
  theoretical predictions based on the DD scenario by Ruiz-Lapuente \&
  Canal (1998, dotted), Yungelson \& Livio (2000, dot-dashed), Greggio
  (2005, solid), and Belczynski et al. (2005, dashed).  The data
  points are the same as those in Fig. \ref{fig:dtd_obs}.  The model
  curves are normalized by the DTD data at $t_{\rm Ia}$ = 11 Gyr, but
  the normalization is arbitrary and only the DTD shapes should be
  compared between the data and models.}
\label{fig:dtd_DD}
\end{figure*}

The DTDs predicted by the four different theoretical models
(Ruiz-Lapuente \& Canal 1998; Yungelson \& Livio 2000; Greggio 2005;
Belczynski et al. 2005) based on the DD scenario are shown in Fig.
\ref{fig:dtd_DD} in comparison with the observed DTD data points.  The
model curves are normalized at the data point of $t_{\rm Ia} = 11$
Gyr, because the DTD behavior at large delay time is especially
important for this scenario from the theoretical point of view (see
below).  However, the normalization should be regarded as a free
parameter and we should compare only DTD shapes between the
observations and the models.  It is impressive that the predictions by
different authors are very close, and in excellent agreement with the
observed DTD at $t_{\rm Ia} \gtrsim$ 0.2 Gyr. It is not surprising
that different authors made similar predictions, because it is a
robust and generic prediction of the DD scenario as argued below.

A common feature for DTD models based on this scenario is that the
delay time is mainly determined by the time $t_{\rm GW}$ from the
formation of a DD binary (i.e., a binary of two white dwarfs) to a
merger after the angular momentum loss by gravitational wave
radiation, especially in the long delay time range of $t_{\rm Ia}
\gtrsim 1$ Gyr.  The general relativity tells us $t_{\rm Ia} \sim
t_{\rm GW} \propto a^4$, where $a$ is the initial separation of the DD
binary.  If the separation distribution is given by $f_{\rm sep}(a)
\propto a^\beta$, the DTD should be
\begin{eqnarray}
f_D(t_{\rm Ia}) \ \propto \ f_{\rm sep}(a) \ \frac{da}{dt_{\rm Ia}}
\ \propto \ t_{\rm Ia}^{-(3-\beta)/4} \ ,
\end{eqnarray}
and hence $\alpha = - (3 - \beta) / 4$.  It is known that the
distribution of initial binary separation $a_0$ at the time of binary
star formation is approximately flat in $\log a_0$, i.e., $f_{\rm sep,
  0} (a_0) \propto a_0^{-1}$ (Abt 1983).  Although there is a
significant change and contraction of binary separations during the
evolution from a binary star formation to a DD binary (e.g., Greggio
2005), it would be reasonable to assume $\beta \sim -1$ also for the
initial separation of DD binaries. Then we expect $\alpha \sim -1$, as
found in the DTD models based on the DD scenario. Though $\beta$ is
rather uncertain and model-dependent, the dependence of $\alpha$ on
$\beta$ is small. Furthermore, the range of the delay time from 0.1 to
10 Gyr corresponds to a range of $a$ only by a factor of 3.2.
Therefore, if $a$ is smoothly distributed in this narrow range, we
expect a power-law DTD with $\alpha \sim -1$ in wide and general
conditions. This is a general result applicable to any merging
phenomena triggered by gravitational wave radiation, such as binary
neutron star mergers (Totani 1997), and this is why the DTD models by
different authors are quite similar to each other, especially at large
delay times. The agreement of the measured DTD with this generic
prediction gives a strong support for the DD scenario.

In Fig. \ref{fig:dtd_DD}, we selected the standard DTD model when a few
models based on the DD scenario are presented by a single author group
[the standard DDS model of Belczynski et al. (2005) and the close-DD
model of Greggio (2005)]. The variation models predict slightly
different value of $\alpha$, but smooth power-law like DTDs are always
common predictions, as expected from the above consideration.

On the other hand, the difference between the DTD predictions by
different authors becomes bigger at short delay time ($t_{\rm Ia}
\lesssim 0.1$ Gyr), where the delay time is dominated by stellar
evolution time scale rather than $t_{\rm GW}$.  There are large
uncertainties in the treatments of stellar evolution in binary
population synthesis, and the discrepancy between some of the DTD models
and the observed data at this short $t_{\rm Ia}$ range is not serious
for the DD scenario.

It is generally assumed that DD binaries are formed through the common
envelope evolution phase. If, in addition, DD binaries are formed also
from intermediate mass binaries through the classical Roche lobe
outflow phase, DD binaries of this channel with large separations may
affect the late time behavior of the DTD in the DD scenario (De Donder
\& Vanbeveren 2003).  In other words, the observed DTD gives some
constraints on the efficiency of DD binary formation via the Roche
lobe outflow phase.

\subsection{the Single-Degenerate Scenario}

\begin{figure*}
\begin{center}
\includegraphics[width=100mm,angle=-90]{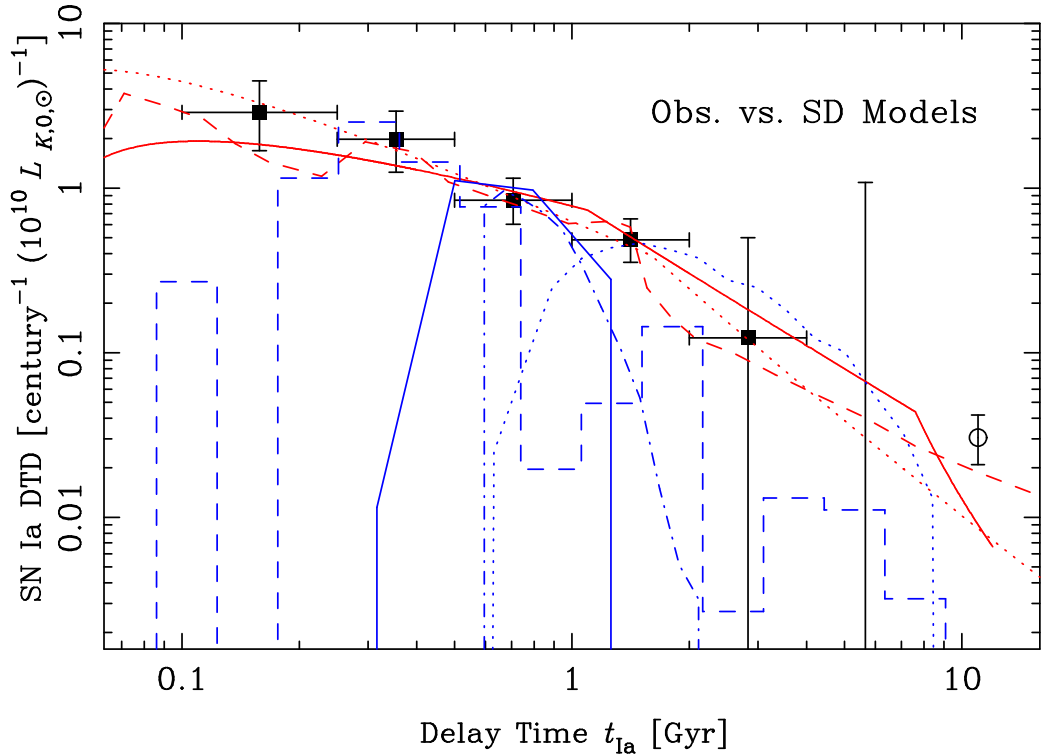}
\end{center}
\caption{The same as Fig. \ref{fig:dtd_DD}, but for a comparison with
  the DTD predictions based on the SD scenario.  The red curves are
  predictions of Greggio (2005, solid curve), Matteucci et al. (2006,
  dotted), and Kobayashi \& Nomoto (2008, dashed), based on analytic
  calculations in which the condition for successful SNe Ia is
  determined by stellar masses in a binary.  The blue curves are
  predictions of Ruiz-Lapuente \& Canal (1998, dotted), Yungelson \&
  Livio (2000, dot-dashed), Belczynski et al. (2005, dashed), and Meng
  et al.  (2008, solid), based on binary stellar population synthesis
  calculations. The theoretical curves are normalized by $\chi^2$
  minimization to the data, and only the DTD functional shapes should
  be compared in this plot.  } \label{fig:dtd_SD}
\end{figure*}

The measured DTD is compared with the theoretical models based on the
SD scenario in Fig. \ref{fig:dtd_SD}. Again, the normalization of the
models is arbitrary, and we should compare only the shape of the DTD
function. Here, the theoretical curves are normalized by $\chi^2$
minimization to the data (our own measurements at 0.1--8 Gyr plus the
11 Gyr data point).  In contrast to the DD scenario, the predictions
by this scenario are quite different depending on different
authors. It is common to this scenario that the delay time is
essentially determined by the main-sequence life time of the secondary
star in a binary, because other time scales (e.g., accretion phase
onto the white dwarf) are much shorter. However, the DTD models shown
in this figure are calculated by considerably different methods with a
wide range of complexity.

The predictions by Matteucci et al. (2006) and the $Z=0.05$ model of
Kobayashi \& Nomoto (2008, hereafter KN08) are close to a simple
power-law, and they are relatively in good agreement with the observed
DTD, compared with other models. However, it should be noted that the
two models are based on the simplest calculations among the models
shown here, though such a simple analytical approach is useful to see
the general behaviors and in applications for e.g., studies of chemical
evolution in galaxies (Kobayashi et al. 1998; Matteucci \& Recchi
2001; Matteucci et al. 2006; KN08).  In these calculations, the
condition for a binary system to evolve to a SN Ia event (i.e., stable
accretion up to the Chandrasekhar mass) is determined by the initial
masses of the primary and secondary stars, regardless of the binary
separation. If the masses are in acceptable ranges, a binary evolves
into a SN Ia with a constant efficiency, and no SN Ia is produced in
other mass ranges. In such a calculation, the DTD shape is determined
by IMF and the distribution of the primary/secondary mass ratio. If
smooth functions are assumed for these, a smooth DTD without a
characteristic scale of $t_{\rm Ia}$ is obtained. However, a constant
SN Ia efficiency within an extended mass range is likely to be an
oversimplification. The SN Ia condition should be determined by the
combination of the stellar masses and binary separation, and it is
natural to expect that there are some characteristic secondary mass
scales preferred for successful SN Ia events, which should appear as
particular time scales in DTD, rather than a power-law like DTD.

In the KN08 model based on the progenitor model of Hachisu et
al. (1996, 2008a), there are two distinct populations of the
main-sequence donor channel (WD+MS) and the red-giant donor channel
(WD+RG) for short and long delay times, respectively.  The relative
abundances of the two are determined empirically by fitting to the
chemical evolution data, rather than by theoretical modeling.
Therefore, the good agreement of the KN08 model with our DTD estimate
may partially be a result of this feedback from other observations,
indicating that our DTD estimate is also consistent with the chemical
evolution data.  Though the WD+MS channel is widely considered as the
promising SD progenitor of SNe Ia, some theoretical calculations
indicate that the SN Ia rate by the WD+RG channel is much lower than
that of the WD+MS channel (Yungelson \& Livio 1998; Han \&
Podsiadlowski 2004; Meng et al. 2008, but see also Hachisu et
al. 2008b).

The SN Ia condition is determined only by stellar masses also in the
standard SD-Chandra model of Greggio (2005), but the mass budget of
accretion up to the Chandrasekhar mass is treated in more detail. An
interesting feature is the sharp drop of DTD at a large delay time of
$\gtrsim 8$ Gyr. SNe Ia with such a large delay time are produced from
binary systems with a small secondary mass.  Such a low mass star has
a small envelope mass that can be used for accretion onto its
companion.  Therefore, the primary stellar mass must be large enough
to ensure that the initial white dwarf mass is sufficiently massive
and the white dwarf successfully grows to the Chandrasekhar
mass. Consequently, the number of binaries that can evolve to SNe Ia
rapidly decreases with decreasing secondary mass.  It seems difficult
to reproduce a single power-law like DTD up to $\sim$10 Gyrs if this
effect is incorporated.

Finally, predictions of the SG-Ch model of Yungelson \& Livio (2000),
the standard SDS model of Belczynski et al. (2005), the CLS model of
Ruiz-Lapuente \& Canal (1998), and the model of Meng et al. (2008)
($\alpha_{\rm CE} = 3$ and $Z = 0.03$) are shown, which are based on
detailed calculations of the binary population synthesis, where the
calculations start from the initial conditions including the primary
and secondary stellar masses and the binary separation. There are a
number of parameters and uncertain physical processes in such
calculations, and different conditions for SNe Ia are assumed by
different authors, leading to vastly different DTD
predictions. However, a clear trend is that some characteristic scales
of the delay time appear in the DTD, making the DTD shape more complex
than the simple analytic models. This is reasonable in realistic
conditions, as argued above, and it is theoretically unlikely that a
simple power-law like DTD is obtained in the framework of the SD
scenario\footnote{After we put the first version of this paper on the
  preprint server (arXiv:0804.0909v1), Hachisu, Kato, \& Nomoto
  (2008b) performed a DTD calculation based on binary population
  synthesis incorporating their progenitor model.  An important
  difference of this calculation from KN08 is that the ratio of the
  two evolutionary channels (WD+MS and WD+RG) is calculated
  theoretically rather than determined by fitting to observed data.
  They found that their DTD prediction is in good agreement with the
  data derived here, indicating that there is a viable model parameter
  space in the SD scenario. On the other hand, each of the two
  channels has a narrow DTD shape similar to the other SD models based on
  binary population synthesis, confirming the trend of the SD
  predictions.  It seems somewhat a fine-tuning that two independent
  components form a featureless power-law DTD as a sum, although it
  could well happen in the nature.  The ratio between the two
  components depends on still uncertain physical processes, such as
  mass-stripping of secondary stars by wind from accreting white
  dwarfs. Further theoretical investigation is important.}.

\section{Conclusions}
\label{section:conclusions}

\subsection{Summary of This Work}
We measured the delay time distribution (DTD) of type Ia supernovae by
using the statistics of the faint variable objects detected in the
systematic variable object survey performed as a part of the
Subaru/XMM-Newton Deep Survey project.

Based on photometric redshift calculations using 9 band
($BVR_ci'z'JK$, 3.6 $\mu$m, and 4.5 $\mu$m) photometries, we selected
16,492 old galaxies from the SXDS data in the redshift range of $0.4
\leq z \leq 1.2$, by requiring that their SED can be fit by a single
starburst and their ages are significantly greater than the star
formation time scales of the bursts.  This selection is essential to
our work, because stellar age is expected to be a good estimator of Ia
delay time in such galaxies. Furthermore, we have two more merits of
this selection: (1) delayed SNe Ia are dominant compared with CC SNe
and prompt SNe Ia, and (2) extinction effect should be small.

We then selected variable objects
associated with these galaxies but having significant offsets from the
nuclei of galaxies, to remove the contamination of AGNs. We found 65
variable objects, whose locations with respect to their host galaxies
closely trace the profiles of galactic light, and hence the majority
of them must be supernovae.  Though we do not have spectroscopic
confirmation of the SN types, we quantitatively demonstrated that the
majority ($\gtrsim$80 \%) of the SN candidates should be SNe Ia, based
on the variability luminosity, redshift distribution, and properties
of the host galaxies. 

Then the DTD in the delay time range of $t_{\rm Ia}$ = 0.1--8 Gyr is
derived by calculating the delay time for each SN Ia candidate from
age and star formation history of host galaxies.  Combined with the
observed SN Ia rate in elliptical galaxies in the local universe, we
derive the SN Ia DTD in a range of $t_{\rm Ia} = $ 0.1--11 Gyr, and
found that it can well be described by a simple power-law, $f_D(t_{\rm
  Ia}) = 0.55^{+0.12}_{-0.11} (t_{\rm Ia}/{1 \ \rm Gyr})^{\alpha} \
{\rm century}^{-1} (10^{10} L_{K, 0, \odot})^{-1}$ with $\alpha =
-1.08^{+0.15}_{-0.15}$.  Here, the DTD function $f_D(t_{\rm Ia})$ is
per unit delay time and per unit mass of a single-burst stellar
population, and $L_{K,0}$ ($K$-band luminosity at an age of 11 Gyr) is
used as an observational estimator of the stellar mass.  We performed
various tests about the systematic uncertainties in this DTD
measurement, but the changes of DTD estimates are not large enough to
change our main conclusions significantly.  We tried a variety of DTD
calculations with different prescriptions as shown in
Fig. \ref{fig:dtd_syst}, but the decreasing trend of $f_D(t_{\rm Ia})$
consistent with $f_D(t_{\rm Ia}) \propto t_{\rm Ia}^{-1}$ at $t_{\rm
  Ia} \sim $ 0.3--3 Gyr is found in all cases, and the data points at
0.5--2 Gyr are especially robust against examined systematic
uncertainties.

The derived DTD at $t_{\rm Ia} \gtrsim$ 0.2 Gyr is in excellent
agreement with the theoretical predictions based on the DD scenario. The
theoretical predictions by different authors are very similar to each
other, and a featureless power-law shape is inevitable consequence of
the general relativity in this scenario. Therefore we consider that the
agreement between the observed and predicted DTDs gives a strong support
to this scenario. It indicates that the major contributor to SNe Ia is
the DD channel for delay times larger than $t_{\rm Ia} \gtrsim 0.1$ Gyr,
although some contribution from other channels cannot be excluded.

On the other hand, the predictions by the competing SD scenario are
vastly different for different authors. Although the predicted DTDs
based on simple analytical approaches have smooth shapes which are
broadly consistent with our measurement, predictions based on more
detailed binary population synthesis calculations show strong peaks in
the DTD shape, which do not fit the observed DTD.  This trend can
naturally be understood; if there is any preferred scale of the
secondary stellar mass for a binary to successfully evolve into a SN
Ia, it should be reflected as a characteristic delay time scale in
this scenario. Therefore we consider that our result does not favor
the SD scenario in general as the major channel to the delayed SN Ia
population ($t_{\rm Ia} \gtrsim$ 0.1 Gyr). However, there are many
degrees of freedom and uncertainties in the theoretical modeling based
on the SD scenario, and it would be premature to reject the SD
scenario simply from our result. Our result should set strong
constraints on the model parameter space if the SD channel is
responsible for the majority of the delayed SNe Ia, and it would be
useful for the future theoretical studies of the SD progenitor
models. It is highly desirable to examine physical effects and
evolutionary paths that are not taken into account in existing models
of binary population synthesis.

\subsection{Discussion and Other Implications}

In the literature, it has occasionally been argued that the SD
scenario is more favored than the DD scenario, but arguments against
the DD scenario are not particularly strong. A merger of two white
dwarfs may result in an accretion induced collapse rather than a SN Ia
(Saio \& Nomoto 1985), but theoretical uncertainties are still quite
large (Piersanti et al. 2003; Yoon et al. 2007).  Many binary systems
are proposed as candidates of the SN Ia progenitor in the SD
framework, compared with the observed number of DD binaries having
total masses larger than the Chandrasekhar mass (Parthasarathy et
al. 2007).  However, DD binaries are difficult to detect and the
statistics of DD binary searches is not sufficient yet to confirm or
reject the DD scenario (Tovmassian et al. 2004; Geier et al. 2007;
Napiwotzki 2007).  There are some implications for the progenitor from
studies of SN Ia remnants (Ruiz-Lapuente et al. 2004; Badenes et
al. 2007; Ihara et al. 2007), or from spectroscopic studies (Leonard
2007; Patat et al. 2007a, b; Simon et al. 2007) and archival
progenitor searches (Voss \& Nelemans 2008; Nelemans et al. 2008;
Roelofs et al. 2008) for nearby SNe Ia.  However, conclusive results
from these methods have not yet been obtained, mainly because of the
limited statistics and theoretical uncertainties.

Another possibility is the SD sub-Chandra scenario (also known as
helium ignitors or edge-lit detonations), in which white dwarfs
explode as SNe Ia before they reach the Chandrasekhar mass.  Since the
required amount of accreting mass is smaller than in the SD Chandra
scenario, binaries with lower secondary masses can more easily evolve
to SNe Ia. This might be useful to overcome the difficulty of the SD
scenario to explain SN Ia rate at $t_{\rm Ia} \gtrsim 10$ Gyr (Greggio
2005).  However, this scenario is currently not popular because
predicted spectra are in serious disagreement with observations
(Hillebrandt \& Niemeyer 2000; Livio 2001).

Recent studies (Scannapieco \& Bildsten 2005; Sullivan et al. 2006;
Aubourg et al. 2007) have tried to model the DTD by two components:
the prompt component proportional to SFR and the delayed component
proportional to stellar mass (i.e., constant $f_D$).  Although our
result does not put a strong constraint on the amount of the prompt
component, a constant $f_D$ seems to be an oversimplification for the
delayed component, since $f_D(t_{\rm Ia})$ at an intermediate delay
time of $t_{\rm Ia} \sim $ 1 Gyr is about 10 times larger than
$f_D({\rm 11 \ Gyr})$. It should be noted that the prompt Ia
population inferred from our data is a considerable fraction ($\sim
20$\%) of all SNe Ia integrated over $t_{\rm Ia} = $ 0--11 Gyr, if we
make a modest assumption that $f_D (t_{\rm Ia})$ at $t_{\rm Ia} < $
0.1 Gyr is constant at the value of $f_D({\rm 0.1 \ Gyr})$. Most of
the observed SN Ia rate dependence on galaxy properties can be
reproduced by a DTD similar to those predicted by the DD scenario
(Greggio 2005; Mannucci et al. 2006).  The observed enhancement of SN
Ia rate in radio galaxies may require an even higher prompt Ia
fraction than that expected from our data or the DD scenario (Della
Valle et al. 2005; Mannucci et al. 2006), if the enhancement is due to
a recent starburst connected to AGN activity.  However, this
interpretation seems to be inconsistent with the fact that no
enhancement of CC SN rate is observed in radio galaxies (Greggio,
Renzini, \& Daddi 2008).  The statistics of this enhancement is still
small (2$\sigma$ level), and it must be confirmed by future
observations.

Recently, Pritchet et al. (2008) reported that DTD proportional to
$t_{\rm Ia}^{-0.5 \pm 0.2}$ is implied based on the SNLS data
(Sullivan et al. 2006), which seems to be inconsistent with our
results. It should be noted that their constraint on DTD is not based
on delay time estimate for each SN Ia, but it is indirectly derived
from the correlation between SN Ia rate per unit galaxy mass (specific
SN Ia rate) and SFR per unit galaxy mass (specific SFR). It is
difficult to estimate how large is the systematic uncertainty in such
an analysis.  Especially, low specific SFR galaxies should be treated
with caution. There should be no tight relation between specific SN Ia
rate versus specific SFR in such galaxies; specific SFR could change
significantly by changing SFR with a fixed stellar mass, but specific
SN Ia rate is hardly affected if it is dominated by old stellar
population.  This should limit the power of this approach to constrain
DTD. It should also be noted that the stellar mass estimates in
Sullivan et al. (2006) are based on five optical bands
($u^*g'r'i'z'$), although near-infrared bands are essential to
reliably estimate stellar masses of galaxies. Another important point
is that we obtained a strong constraint on the slope index $\alpha$ by
supportively using the SN Ia rate in nearby elliptical galaxies.  If
we fit only our own DTD data, we obtain $\alpha =
-0.92^{+0.30}_{-0.27}$.  It is important to combine the nearby rate
data (large $t_{\rm Ia}$) and high-$z$ data (small $t_{\rm Ia}$) to
get a strong constraint on the DTD shape in a wide range of the Ia
delay time.

\bigskip

We would like to thank K. Belczynski, R. Canal, L. Greggio,
F. Matteucci, X. Meng, P. Ruiz-Lapuente, and L. R. Yungelson for
providing numerical data of their DTD models. This work is based in
part on the observations made with the Subaru Telescope operated by
the National Astronomical Observatory, the United Kingdom Infrared
Telescope operated by the Joint Astronomy Centre on behalf of the
Science and Technology Facilities Council of the U.K., and the Spitzer
Space Telescope operated by the Jet Propulsion Laboratory, California
Institute of Technology under a contract with NASA of the U.S.A. A
part of the optical imaging and spectroscopic data were obtained as a
part of the Supernova Cosmology Project. This work was supported in
part by the Grant-in-Aid for Scientific Research (19740099, 19035005),
for the 21st Century COE Program ``Center for Diversity and
Universality in Physics'', and for the Global COE Program "The Next
Generation of Physics, Spun from Universality and Emergence" from the
Ministry of Education, Culture, Sports, Science and Technology (MEXT).
This work was also supported in part by the Core-to-Core Program
``International Research Network for Dark Energy'' and the Japan-USA
Bilateral Program of the Japan Society for Promotion of Science
(JSPS).



\end{document}